\documentclass[letterpaper,11pt]{report}

\usepackage{fullpage}
\usepackage{verbatim}
\usepackage{cite}
\usepackage{setspace}
\usepackage{fancyhdr}
\usepackage{listings, xcolor, soul}
\usepackage{scrextend}
\usepackage[small]{caption}
\usepackage{graphics}
\usepackage{color}

\usepackage{hyperref}

\usepackage[dvips]{graphicx}

\usepackage{amsmath}
\usepackage{booktabs,multirow}
\usepackage{hhline}
\usepackage{caption}
\usepackage{subfigure}
\usepackage{capt-of}
\usepackage{multirow}
\usepackage{array}
\usepackage{multicol}
\def\title{Mytitle}

\begin{document}

\def\addrone{Your address}
\def\addrtwo{Your city}

\def\degree{M.Tech. in Computer Science with Specialization in Data Engineering}

\def\submissiondate{June 06, 2014}

\def\supervisorone{Dr. Ponnurangam Kumarguru}

\def\supervisortwo{Dr. Vinayak Naik}

\def\supervisorthree{Dr. Ayesha Choudhary}

\thispagestyle{empty}

\begin{center}

{\LARGE \bf {Twitter and Polls: Analyzing and estimating political orientation of Twitter users in India General \#Elections2014  }

 }  
 \vspace{1in}
 
 {\Large{Student Name: Abhishek Bhola}} \\  
 \vspace{.1in} 
 IIIT-D-MTech-CS-DE-12-MT12031 \\

 June 06, 2014 \\
  
    \vspace{.6in}

  \vspace{.5in}

{Indraprastha Institute of Information Technology\\
New Delhi}

\vspace{.8in}  {\underline{Thesis Committee} \\ \supervisorone                ~(Chair) \\ \supervisortwo \\ \supervisorthree} \\ \vspace{.35in}

\vspace{.6in} {Submitted in partial fulfillment of the requirements \\for the Degree of M.Tech. in Computer Science, \\ with specialization in Data Engineering}

\vspace{.8in}

\copyright 2014 Abhishek Bhola  \\ All rights reserved \\
\vspace{.8in}

\end{center}

\newpage

\pagestyle{empty}
\vspace*{7.1in} 

\newpage

\begin{center}
\section*{Certificate}\label{section:certificate}
\end{center}
This is to certify that the thesis titled \textbf{``Twitter and Polls: Analyzing and estimating political orientation of Twitter users in India General \#Elections2014"} submitted by \textbf{Abhishek Bhola} for the partial fulfillment of the requirements for the degree of \emph{Master of Technology} in \emph{Computer Science \& Engineering} is a record of the bonafide work carried out by him under my guidance and supervision in the Data Engineering group at Indraprastha Institute of Information Technology, Delhi. This work has not been submitted anywhere else for the reward of any other degree. \\ \vspace{0.5in}

\textbf{Professor Ponnurangam Kumarguru}\\
\textbf{Indraprastha Institute of Information Technology, New Delhi}

\begin{abstract}

This year (2014) in the month of May, the tenure of the 15th Lok Sabha was to end and the elections to the 543 parliamentary seats were to be held. With 813 million registered voters, out of which a 100 million were first time voters, India is the world's largest democracy. A whooping \$5 billion were spent on these elections, which made us stand second only to the US Presidential elections (\$7 billion) in terms of money spent. The different phases of elections were held on 9 days spanning over the months of April and May, making it the most elaborate exercise to choose the Prime Minister of India. Swelling number of Internet users and Online Social Media (OSM) users turned these unconventional media platforms into key medium in these elections; that could effect 3-4\% of urban population votes as per a report of IAMAI (Internet \& Mobile Association of India). Political parties making use of Google+ Hangout to interact with people and party workers, posting campaigning photos on Instagram and videos on YouTube, debating on Twitter and Facebook were strong indicators of the impact of the OSM on the General Elections 2014. With hardly any political leader or party not having his account on the micro blogging site Twitter and the surge in the political conversations on Twitter, inspired us to take the opportunity to study and analyze this huge ocean of elections data. Our count of tweets related to elections from September 2013 to May 2014, collected with the help of Twitter's Streaming API was close to 18.21 million.

We analyzed the complete dataset to find interesting patterns in it and also to verify if the trivial things were also evident in the data collected. We found that the activity on Twitter peaked during important events related to elections. It was evident from our data that the political behavior of the politicians affected their followers count and thus popularity on Twitter. Yet another aim of our work was to find an efficient way to classify the political orientation of the users on Twitter. To accomplish this task, we used four different techniques: two were based on the content of the tweets made by the user, one on the user based features and another one based on community detection algorithm on the retweet and user mention networks. We found that the community detection algorithm worked best with an efficiency of more than 80\%. It was also seen that the content based methods did not fare well in the classification results. With an aim to monitor the daily incoming data, we built a portal to show the analysis of the tweets of the last 24 hours.\footnote{\url{http://bheem.iiitd.edu.in/IndiaElections}} This portal analyzed the tweets to find the most trending topics, hashtags, the kind of sentiments received by the parties, location of the tweets and also monitored the popularity of various political leaders and their parties' accounts on Twitter.  To the best of our knowledge, this is the first academic pursuit to analyze the elections data and classify the users in the India General Elections 2014.

\end{abstract}

\newpage
\pagestyle{empty}

\newpage

\section*{Acknowledgments}\label{section:acknowledgments}
\pagestyle{plain}
\pagenumbering{roman}

I would like to express my special appreciation and thanks to my advisor, \textit{Dr. PK}, you have been a tremendous mentor for me. You continually and persuasively conveyed a spirit of adventure in the work. Without your encouragement and advise my thesis could not have been materialized. I would also like to thank my committee members \textit{Dr. Vinayak Naik} and \textit{Dr. Ayesha Choudhary} for serving as my committee members even at hardship. 

My intellectual debt is to my shepherd \textit{Anupama Aggarwal} whose insightful comments and suggestions helped me at every stage. A special thanks to Precog members, especially \textit{Paridhi Jain} whose constructive comments and warm suggestions always helped me. I need to thank the members of CERC at IIIT-Delhi for providing me a platform to present and discuss my work. 

This work could not have been possible if the annotators did not do the data annotations for me; \textit{Ritika, Megha, Aayushee, Niharika, Prateek, Mayank, Geetanjali, Srishti and Prachi} my deepest gratitude to all of you. And finally no words are sufficient enough to thank my parents, my backbone.
\newpage

\tableofcontents
\listoffigures 
\listoftables

\newpage

\newpage

\newpage
\mbox{}


\chapter{Introduction}\label{chapter:introduction}
\pagenumbering{arabic}
\setcounter{page}{1}
\onehalfspacing

During the year 2014, the world's largest democracy, India witnessed elections for the 16th Lok Sabha. The general elections is the biggest exercise to form the government at the center and thus elect the Prime Minister of India. These elections are held every five years when the incumbent government is nearing its completion. The India General Elections were held during the months of April and May for 543 Parliamentary constituencies all over India. For a party to form the government, it must win atleast 272 seats or prove its majority with the help of alliances. The responsibility of conducting elections in a fair manner lies with the Election Commission of India (ECI). The total number of registered voters with the Election Commission of India are 814.5 million, in comparison to 713 million in 2009.\footnote{\label{note1}Press note released by the ECI dated 05.03.2014} This marks a 100 million rise in the number of newly registered voters.

\begin{figure}[!ht]
\centering
\includegraphics[scale=.4]{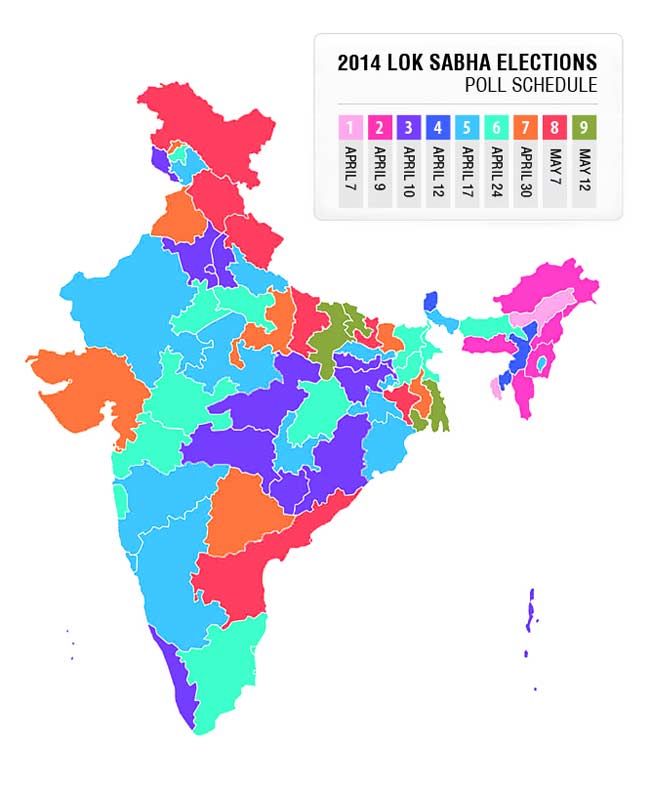}
\caption[Map of India showing the nine phases of elections]{Nine phases in which elections were held over 543 constituencies\tiny{\footnotemark}}
\label{fig:ff1}
\end{figure}

The number of polling stations across the country was 0.9 million with 1.72 million control units and 1.87 million ballot units.\footref{note1} The amount of money spent by the parties in these elections was around \$5 billion, a sum second only to the US Presidential elections which stands first with \$7 billion.\footnote{http://www.ibtimes.com/indias-2014-election-cost-5-billion-second-only-price-tag-2012-us-presidential-election-1570668} The elections to the 543 Parliamentary constituencies was held in 9 phases spread over the months of April and May; the first one held on 07th April 2014 and the ninth phase held on 12th May 2014. The fifth phase was the biggest one when elections at 122 constituencies were held simultaneously and the first phase was the smallest with only 6 constituencies having elections on that day. A map depicting the nine phases can be found in the Figure \ref{fig:ff1}. 

\footnotetext{Source: \url{http://bit.ly/1k6yUgP}}
The number of parties registered with the ECI are $1,616$. This number consists of $6$ national parties, $47$ state parties and others as registered unrecognized parties. The elections of 2014 had a battle between few major parties, namely the Indian National Congress (INC) or commonly known as Congress, the Bhartiya Janta Party (BJP), the Aam Aadmi Party (AAP) and some other national parties like the Samajwadi Party (SP), the Bahujan Samaj Party (BSP) etc. forming the third front. The candidates who remained in news for most of the times were the Prime Ministerial candidates of these parties, like Narendra Modi from  the BJP, Arvind Kejriwal from AAP and Rahul Gandhi from Congress, although Rahul Gandhi was never officially declared as the PM candidate.

\section{Elections and Social Media}
There was a significant change in the General Elections 2014 from the General Elections 2009; this was the change in the role played by the social media during the elections. It has been observed worldwide that the democracies have been engaging in dialogues with the public over the social media \cite{hong2011does}. As per Internet and Mobile Association of India (IAMAI) the Internet users in India are expected to be $243$ million by June 2014, marking a $28$\% growth from June 2013.\footnote{\url{http://bit.ly/1ry6yRk}}  There are a variety of social media platforms available, e.g., Facebook, Twitter, Pinterest, Instagram, Tumblr, Flickr, Google+, LinkedIn to name a few. The people are increasingly using these platforms to express their views on different topics. India has been at the forefront of the growth stories in the number of users of these social media platforms. The Facebook being the most popular has $114.8$ million Indian users.\footnote{\url{http://tech.firstpost.com/news-analysis/india-poised-to-overtake-us-as-facebooks-no-1-market-215873.html}} Twitter ranks second in the number of users with $33$ million users from India.

Call it an influence of the US Presidential Elections or not, but the social media platforms were used by the parties for their campaigns. Almost all the major parties made their presence felt on the social media with the official accounts and verified pages of their leaders and the parties. As per the media reports, professional help was taken by the parties to improve the image of their party and leaders on the social networking sites. As per a study of IAMAI, there were some $160$ odd constituencies out of $543$ that were likely to be influenced by these new media and also sway $3-4$\% of votes.\footnote{\url{http://forbesindia.com/article/real-issue/social-media-limited-but-liked-in-indian-elections/37550/0}} Some parties like the BJP and Congress made the use of Google+ Hangout to conduct meetings with people.\footnote{\url{http://indiatoday.intoday.in/story/rahul-gandhi-interact-with-party-workers-on-google-hangout/1/349706.html}} Candidates used the medium of televised interviews on Facebook and WhatsApp messaging service to connect to the millions of urban voters. Campaign videos were uploaded on YouTube by several parties. A lot of campaigning ads were seen on Facebook during the elections, while all the photographs of Rahul Gandhi's rallies were uploaded on an Instagram account.\footnote{http://instagram.com/inc\_india} Google did its bit by introducing the Google Election hub to empower voters and by giving more information about the elections and the candidates.\footnote{\url{http://www.google.co.in/elections/}}

\section{Research Motivation}
Out of all the social networking platforms, Twitter is lauded by academicians, journalists and politicians for its potential political value. Many politicians make use of this micro blogging site to express themselves in the limit of 140 characters. Twitter has had the provision of verified accounts since long and therefore the communication coming from these accounts looks more authentic and is successful in creating more buzz than the other social media platforms. 

General Elections in India is the biggest democratic exercise in 5 years. And this time with social media as an added factor, it generated a bulk of data everywhere. In literature a lot of studies have been done on the data related to elections on Twitter. A majority of them have been on the U.S Presidential Elections where there are two main candidates who battle it out. There are $50$ million Twitter users in the United States and use English as their medium to communicate on Twitter.\footnote{\url{http://articles.latimes.com/2013/nov/21/news/la-sh-twitter-users-graphics-20131121}} But in India, the scenario is different with as many as $1,616$ parties and nearly $8,000$ candidates. The population in India using Twitter is microscopically small when compared to the actual voting population. With people talking about so many parties and candidates and that too in transliterated messages, the problem of identifying their political inclination becomes wider and interesting. All these factors along with sheer amount of data motivated us to look into the data and find interesting patterns in it and if it was in sync with the real time events as they were happening. 

\section{Research Aim}
With all the real time data with us, we aimed to relate some semantics with the data. We wanted to analyze the data and find out if there were any patterns. We aimed to find out what were the words and hashtags that were trending the most and who was at the center of the network graphs. We wanted to see how many unique users we had and what was the popularity on different parameters for different politicians. Yet another aim of our work was to find out the political orientation of the users on Twitter. To look at which party did the person support or was against. Since the three major parties that were mainly in news were AAP, BJP and Congress, we tried to classify each user into ProAAP, ProBJP or ProCongress categories. And similarly for the anti views also, we tried to categorize the users as AntiAAP, AntiBJP or AntiCongress.

Since a lot of data was coming in every day, we felt the need to have a system which could generate daily analysis for us. We wanted to be able to see the tweets related to elections as and when they come. There was a need to keep a track of the top notch politicians' twitter accounts and how their activities correlated with their popularity on twitter. 

So we can broadly specify the aims of our work as follows
\begin{enumerate}
\item To analyze and draw meaningful inferences from the collection of tweets collected over the entire duration of elections
\item To check the feasibility of development of a classification model to identify the political orientation of the twitter users based on the tweet content and other user based features.
\item To develop a system to analyze and monitor the election related tweets on daily basis.
\end{enumerate}

\chapter{Related Work and Research Contribution}
\section{Related Work}
Twitter has served as a platform for information dissemination, banter, breaking news, spreading rumors and many other purposes. This provides good opportunities to researchers to study the real-time events, like Sakaki et. al. did earthquake detection on the basis of tweets \cite{sakaki2010earthquake} and Aramaki et. al. used it for influenza detection \cite{aramaki2011twitter}. These real-time events can also be studied for the credibility in the tweets as done by Gupta et. al. \cite{gupta2012credibility}. One such real time event that has always made news is Elections. 

\subsection{Prediction of Elections Using Twitter}
Tumasjan et. al. \cite{tumasjan2010predicting} who analyzed  100,000 tweets, in their paper claimed that a mere mention or volume analysis of the tweets related to elections was enough to predict the results. They also said that the people's sentiments in real world are closely related to the tweets' sentiments. Till date, this paper has had the most successful attempt in election predictions. The first attempt at mood detection is known to be done by Bollen et. al. \cite{bollen2011modeling}, where they classified the tweets to be belonging to 6 different categories of moods namely tension, depression, anger, vigor, fatigue and confusion. Jungherr et. al. \cite{jungherr2012pirate} however challenged the results of Tumasjan et. al \cite{tumasjan2010predicting} by saying that the results were time dependent and that all the parties were not taken into account. A paper in counter  response was also published by Tumasjan et. al.\cite{tumasjan2011election} in which they toned down their claim.

We then had papers that started raising doubts over the predictive powers of Twitter for elections. First in this series was the paper by Metaxas et. al. \cite{metaxas2011not}. They concluded the success in prediction of elections as a chance and cautioned that one must look at the demographics of the population before predicting elections. Yet another doubt that was raised on the prediction concerned the credibility of tweets by Castillo et. al. \cite{castillo2011information} and on the users tweeting about elections by Mustafaraj et. al \cite{mustafaraj2011vocal}. 

\subsection{Election Data Analysis}
Skoric et. al. \cite{skoric2012tweets} in their research studied the elections data during the Singapore General Elections. They found out that though the predictive power of Twitter for elections can not be claimed to be as good as in the Germany elections by Tumasjan et. al. \cite{tumasjan2010predicting}, but it is still better than chance. Their mean absolute error was higher than that of previous studies and concluded that Twitter is indicative of the public opinion. Gayo-Avello \cite{gayo2011don} in his paper studied the US Presidential Elections 2008 and shows how analysis of twitter data failed to predict Obama's win even in Texas. He claimed that the twitter data is biased and can not be used as a representative sample. He also challenged the sentiment analysis used in the earlier papers.  

As far as India General Elections 2014 were concerned, twitter data was also analyzed by Simplify360.\footnote{\url{http://simplify360.com/politics/IndianElection2014}} They prepared both short summary as well as a detailed report on the elections data. They prepared a Simplify360 Social Index (SSI) to calculate the popularity of the politicians. Awareness, spread, prominence and favorability were 4 parameters they used for the calculation of their index. The analysis by the NExT Center at the National University of Singapore was a weekly analysis.\footnote{\url{http://www.nextcenter.org/india2014/}}  They would find out the statistics about the three major parties AAP, BJP and Congress from the weekly data and also report the political events that took place that week. Their last section had some political reviews about the three main candidates. Kno.e.sis, a research group at Wright State University also analyzed India General Elections 2014 with the help of Twitris+, a semantic social web application.\footnote{\url{http://analysis.knoesis.org/indiaelection/}} The portal showed the hopefulness for the three major parties. The hopefulness was calculated taking the number of mentions and sentiments of the tweets as parameters. Apart from this, there were several portals by news media houses that showed some portion about the kind of activities going on on Online Social Media (OSM), for e.g., the pages by IBN and  TOI .\footnote{\url{http://ibnlive.in.com/general-elections-2014/}} \footnote{\url{http://timesofindia.indiatimes.com/lok-sabha-elections-2014/elections2014.cms}}

\subsection{Political Orientation Prediction}
Though prediction of election results with Twitter has been a beaten problem, not many attempts have been made at the classification of the political orientation of the users. Conover et. al. \cite{conover2011predicting} studied the 2010 U.S. midterm elections data and predicted the orientation of the twitter users with an efficiency of 95\%. They used the latent semantic analysis for classifying the content of the users' tweets and found high correlation values between the content and the political alignment of the users.

Golbeck et.al. \cite{golbeck2011computing} tried to find the political preferences of the followers of popular news media's accounts on Twitter. They calculated a political preference score (P Score) for all the users based on the accounts they were following. It was suggested that based on the calculation of the audiences' preferences, the leaning of the respective media houses can also be found out. Cohen et. al. \cite{cohen2013classifying} in their work challenged the work of Conover et. al. \cite{conover2011predicting}. They said that the classifier developed by them would not work if it was used for a set of users who were not very politically active on Twitter. They divided the set of users into political figures, politically active and moderate users and compared it with the Conover et. al. data set and found the efficiency of the classifier to be lower in their case.

\section{Research Contribution}
Let us highlight the main contributions of our work.
\begin{itemize}
\item We attempted to classify the political orientation of the Twitter users from India. Due to the variety of political parties and leaders and use of different languages, this work became even more challenging.

\item We analyzed the political orientation of Twitter users for not only the `Pro' category, but also for the `Anti' category. 

\item We showed that the network based on retweets and user mentions works the best for classification in the Indian scenario.

\item After analyzing the data over 5 months, we showed that the weekdays saw the maximum activity on Twitter related to politics.

\item We showed that BJP and its Prime Ministerial candidate Narendra Modi were the highest gainers in the fields of mentions and popularity on Twitter.

\end{itemize}
 
\chapter{Data Collection and Methodology}
\section{Data Collection}
We collected data from Twitter with the help of Twitter API. We made use of both Twitter REST API v$1.1$ as well as Streaming API for different kinds of data collections. The different types of data collected have been discussed in the following subsections.
\begin{figure}[!ht]
\centering
\includegraphics[scale=.8]{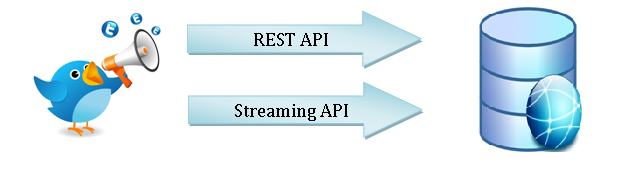}
\caption[Data collection from Twitter with the help of APIs]{Data from twitter is collected via REST and Streaming APIs and stored in our databases}
\label{fig:ff2}
\end{figure}

\subsection{Collecting Election Related Tweets}
We started collecting tweets related to Elections since September 25, 2013. We used the Twitter's Streaming API for the collection of tweets. The method statuses/filter returns the public statuses containing one or more filter predicates.\footnote{https://dev.twitter.com/docs/api/1/post/statuses/filter} These predicates were a lists of keywords related to elections. These lists were manually prepared by us and we tried to be as exhaustive as possible. A few keywords used for tracking tweets is shown in the following Figure \ref{fig:ff3}.

\begin{figure}[!ht]
\centering
\includegraphics[scale=.2]{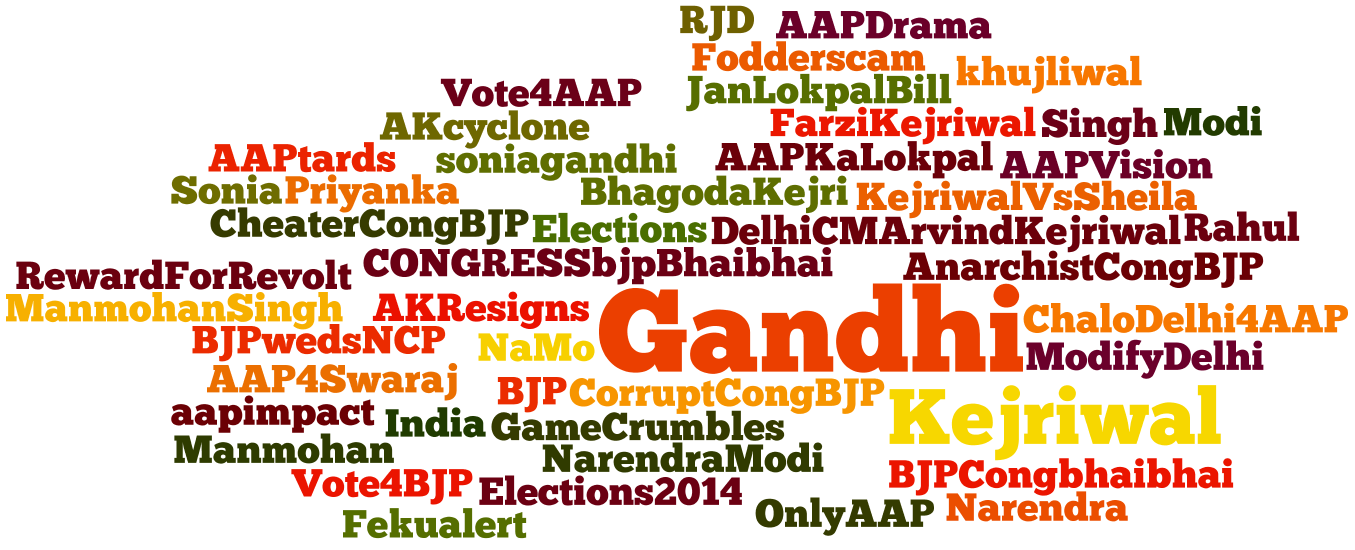}
\caption[Word cloud of keywords used for collecting tweets]{Word cloud of keywords used for collecting tweets from Streaming API\tiny{\footnotemark}}
\label{fig:ff3}
\end{figure}
\footnotetext{Word cloud generated using www.wordle.net}

With this method of tweet collection, we could collect \textbf{18.21 million} tweets till May 16, 2014. As per a report there had been over \textbf{49 million} tweets till April end and there was a $600$\% rise in the number of election related tweets.\footnote{\url{http://bit.ly/1hukpDE}} However, the exact date of start of collection of these tweets and the keywords used were not known. But assuming the collection started from January, we can compare our collection of \textbf{9.83 million} tweets from January to April to the official figures. 

\subsection{Collecting Data About Trending Politicians}\label{3.1.2}
We wanted to keep a track of the twitter accounts of the most trending politicians and the official accounts of their parties, if any. We used the twitter.look\_up method of the Streaming API to collect this data. This method returned the JSON object for the users and provided information such as screen\_name, location, profile\_image\_url, followers\_count, friends\_count, statuses\_count and many other useful pieces of information about that profile. 

We handpicked 130 legitimate twitter profiles of some important politicians and parties. This list included those profiles which were actively using twitter or were of extreme importance in the national politics. We tried to prepare the list as exhaustively as we could. Some famous names in the list were Narendra Modi, Arvind Kejriwal, Shashi Tharoor, Sushma Swaraj, Manish Sisodia, Arun Jaitely etc. The official accounts of parties like BJP, AAP, Trinamool Congress and Congress were also included in this list. Klout score for these profiles was calculated on daily basis and displayed on our system.

\subsection{Collecting Data for Identification of Political Orientation}\label{datacollection}
The third aim of our work required us to create a true positive data set of users with their both `Pro' and `Anti' political orientations known. Since Twitter does not have any provision for mentioning the political orientation of a user in his account details, we had to get it done with the help of our data annotators. We picked 1,000 random unique users from our database who were tweeting about elections and distributed it among the data annotators for annotations. Apart from getting the political orientation of the users, we extracted the rest of the information, viz., the screen\_name, location, followers\_counts, friends\_count etc. from the twitter REST API v1.1. The exact process of data annotations and their results have been discussed in Section \ref{methodology}. This data set was used for developing the classifiers.

\section{Methodology}
\subsection{Methodology for Data Analysis}
The data collection had been on since September 2013. Apart from the daily analysis, we wanted to analyze the overall collection of data to see if there were any interesting patterns in them. As per the definition of data mining, we wanted to be able to give it a structure that is more understandable. To this effect, we used various tools like MS-Excel and `Matplotlib 1.1.1' library of Python to plot the graphs and draw information from the data.

With the help of the graphs and other infographics, we wanted to be able to answer some of the research questions. We developed graphs to monitor the daily and hourly volume analysis. We wanted to find out that which parties got maximum mentions over the period of time. Who were the people who were mentioned the most and how did the popularity of some leaders rise over the period of time were some of the questions we tried answering. We then related all these graphs with the political time line and drew certain inferences that have been discussed in the Chapter \ref{chp5}. 

\subsection{Methodology for identifying political orientation}\label{methodology}
As discussed in Section \ref{datacollection}, we aimed to create a true positive data set of 1000 users for whom the political orientation was known to us. So we decided to get the opinion of human annotators for this task. These annotators were PhD and Masters scholars from IIIT Delhi and IIT Delhi. Each annotator was provided with a set of links of 250 twitter user profiles along with the following instructions.

\begin{lstlisting}
1. Please visit the profile of the twitter user given in the spreadsheet.
2. Look at the tweets of the user. You need not scroll down more than twice.
3. Looking at the political tweets, i.e the tweets related to Indian politics/ elections made by the user, you need to decide 2 things:

	i.  Which party (AAP, BJP or CONG) is the user supporting? Select the name of the party in the PRO column. If you can't make it out from the tweets, kindly write CAN'T SAY.
	ii. Which party (AAP, BJP or CONG) is the user against? Select the name of the party in the ANTI column. If you can't make it out from the tweets, kindly write CAN'T SAY.
	
Please note that you cannot fill more than one party for each of the PRO and ANTI columns. If you find that the user is supporting more than one, please mention CAN'T SAY.
\end{lstlisting}

To avoid any kind of bias each user was annotated by three different annotators. The Pro and Anti classes for the users were decided only if it was agreed upon by majority of the annotators. For each of the Pro and Anti classes, the annotators had the option of labeling the user as Pro / Anti to AAP, BJP, CONG or CAN'T SAY.

As the data annotators were analyzing the users to decide their political orientation, we collected the recent tweets of the users from their time line. We collected 200 tweets of each user from their timeline. Apart from the tweets, we gathered other information about the users such as location, number of friends and followers, number of tweets etc. 

We shall now discuss the results of annotations. Let us look at the 1st set of $250$ users and what was the status of inter-annotator agreement for them. The confusion matrix for the first set of users is shown in the Table \ref{tab:tableA}.

\begin{table}[!h]
\begin{center}
    \begin{tabular}{|l|l|l|l|l|}
    \hline
    \textbf{Party}           & \textbf{AAP}	& \textbf{BJP}	& \textbf{CONG}	& \textbf{CAN'T SAY} \\ \hline
    \textbf{AAP}      		& 18	& 4		& 0  	& 3     	\\ \hline
    \textbf{BJP}           	& 6		& 76	& 1     & 21  		\\ \hline
    \textbf{CONG} 			& 0     & 4     & 2     & 3 		\\ \hline
    \textbf{CAN'T SAY}	   	& 11    & 11    & 4     & 86  		\\ \hline
    \end{tabular}
\caption{Annotations for 1st set of 250 users by two annotators for the `Pro' category.}\label{tab:tableA}
\end{center}
\end{table}

The annotators agreed over 182/250 users for the first set of users. This is the observed agreement of the two annotators. However, to find out the \textbf{reliability} of annotations, the observed agreement was not a sufficient parameter \cite{krenn2004determining}, because it did not take into consideration the agreement that was due to \textit{chance} and we needed to find the chance agreement. We did so by calculating the \textbf{Cohen's kappa coefficient}\cite{carletta1996assessing} :
\begin{equation}
\kappa = \frac{Pr(a)-Pr(e)}{1-Pr(e)}
\end{equation}
where Pr(a) is the observed agreement and Pr(e) is the hypothetical probability of chance agreement.\footnote{\url{http://en.wikipedia.org/wiki/Cohen's_kappa}} The observed agreement for this set has been stated as $183/250= 0.732$. The hypothetical probability was calculated as $0.375$. Thus the chance agreement was calculated as \textbf{0.571}. Since there are more than 2 annotators, we repeated the same procedure to find the Cohen's kappa coefficient for all 3 possible pairs and averaged it and found it be \textbf{0.592}.

Let us now take a look at the same set of 250 users annotated by the same set of annotators for the `Anti' category.

\begin{table}[!h]
\begin{center}
    \begin{tabular}{|l|l|l|l|l|}
    \hline
    \textbf{Party}           & \textbf{AAP}	& \textbf{BJP}	& \textbf{CONG}	& \textbf{CAN'T SAY} \\ \hline
    \textbf{AAP}      		& 20	& 1		& 0  	& 9     	\\ \hline
    \textbf{BJP}           	& 2		& 11	& 2     & 8  		\\ \hline
    \textbf{CONG} 			& 4     & 0     & 9     & 23 		\\ \hline
    \textbf{CAN'T SAY}	   	& 21    & 6     & 6     & 128  		\\ \hline
    \end{tabular}
\caption[Annotations for the 1st set of 250 users by two annotators for the `Anti' category]{Annotations for the 1st set of 250 users by two annotators for the `Anti' category.}\label{tab:tableB}
\end{center}
\end{table}

The observed agreement among the annotators for `Anti' category was 168/250 = 0.672. The hypothetical probability for this case was 0.471. We can thus calculated the Cohen's Kappa coefficient which came out to be \textbf{0.379}.

The observed agreement and hypothetical probability values for all the 4 sets of 250 users and thus 1,000 users is given in the following Table \ref{tab:tableC}.

\begin{table}[!h]
\begin{center}
	\begin{tabular}{|c|p{1cm}|p{1cm}|p{1cm}|p{1cm}|p{1cm}|p{1cm}|}
	\hline
	\multirow{2}{*}{\textbf{Set of Users}} &\multicolumn{2}{p{2.5cm}|}{\textbf{Observed Agreement}} &\multicolumn{2}{p{2.5cm}|}{\textbf{Hypothetical Agreement}} &\multicolumn{2}{p{2.5cm}|}{\textbf{Kappa Coefficient}}\\ 
	\hhline{~------}
 	& Pro & Anti  & Pro   & Anti  & Pro   & Anti	\\\hline
	1st Set of 250 users &0.732 & 0.672 & 0.375 & 0.471 & 0.571 & 0.379	\\ \hline
	2nd Set of 250 Users &0.736 & 0.641 & 0.372 & 0.432 & 0.579 & 0.367	\\ \hline
	3rd Set of 250 Users &0.752 & 0.684 & 0.378 & 0.454 & 0.601 & 0.421	\\ \hline
	4th Set of 250 Users &0.725 & 0.549 & 0.371 & 0.429 & 0.562 & 0.288	\\ \hline
	\end{tabular}
\caption[Agreement values between annotators for all 1,000 users]{Agreement values between annotators for all 1000 users.}\label{tab:tableC}
\end{center}
\end{table}

The result after the inter annotator agreement for all the users can be found in Table \ref{tab:tableD}. We can see that out of 1,000 users that we got annotated, we got 452 such users out of 1,000 for whom the `Pro' category of their political orientation could be decided with agreement between all 3 annotators. Similarly for the `Anti' political orientation, we could find 327 such users for whom the class was not CAN'T SAY. 

\begin{table}[!h]
\begin{center}
    \begin{tabular}{|l|l|l|}
    \hline
    \textbf{Party}          & \textbf{Pro}	& \textbf{Anti}	\\ \hline
    \textbf{AAP}      		& 133	        & 205			\\ \hline
    \textbf{BJP}           	& 447		    & 85	 		\\ \hline
    \textbf{CONG} 			& 33            & 135      		\\ \hline
    \textbf{CAN'T SAY}	   	& 387           & 575     		\\ \hline
    \end{tabular}
\caption[Annotations results]{Annotations results}\label{tab:tableD}
\end{center}
\end{table}


One observation that we could make from this data set was that the number of users Pro to BJP was the highest and the number of users Anti to AAP were highest. Once this true positive data set was created, it was used for developing the classifiers for both Pro and Anti categories. The different methods of classification and their efficiency have been discussed in Chapter \ref{chp6}.

\subsection{Methodology for System Design}\label{3.2.1}
To accomplish the task of monitoring the data on daily basis, we wanted to have a system that could cater to this need. The Elections portal we developed showed the data we collected and the analysis were generated with the help of automated scripts.  

The aims of this portal were to have some analysis on the tweets viz., to see where were these tweets coming from and to visualize them on the map. Another requirement was to find out the the words that were trending. A word cloud was needed to visualize the most trending words. Hashtags are an integral part of the tweets and are used by people to follow and contribute to a particular topic \cite{bruns2011use}. Therefore a graph was required to show the per hour frequency of these hashtags. Apart from these analysis, it was required to look at the retweet and user mention network of the users who tweeted in the last 24 hours. To analyze the sentiments attached with the tweets of the three major parties, i.e., AAP, BJP and Congress, a sentiments tab in the portal was desired.

The front end of the portal was developed using PHP and Javascripts, whereas the data stored in the databases was updated and retrieved using MySQL.

\chapter{Data Analysis}\label{chp5}
In this chapter we will discuss about the first aim of our work, i.e., to analyze the overall data collected over more than 5 months. We shall try to answer some of the research questions from the data that we have. The different type of analysis have been mentioned under different sections.

\section{Volume Analysis}
Based on our set of keywords for collecting the tweets, the data that we could collect from September 25, 2013 to May 16, 2014 was 18,214,947 tweets. Gayo-Avello in his paper on the analysis of twitter data during the US Presidential Elections 2008 claimed that the "conversation" about elections grew as the campaign progressed and elections came nearer \cite{gayo2011don}. To verify this claim we did an analysis of the per day volume of the tweets and compared it with the timeline of the elections.

\begin{figure}[!ht]
\centering
\includegraphics[scale=.5]{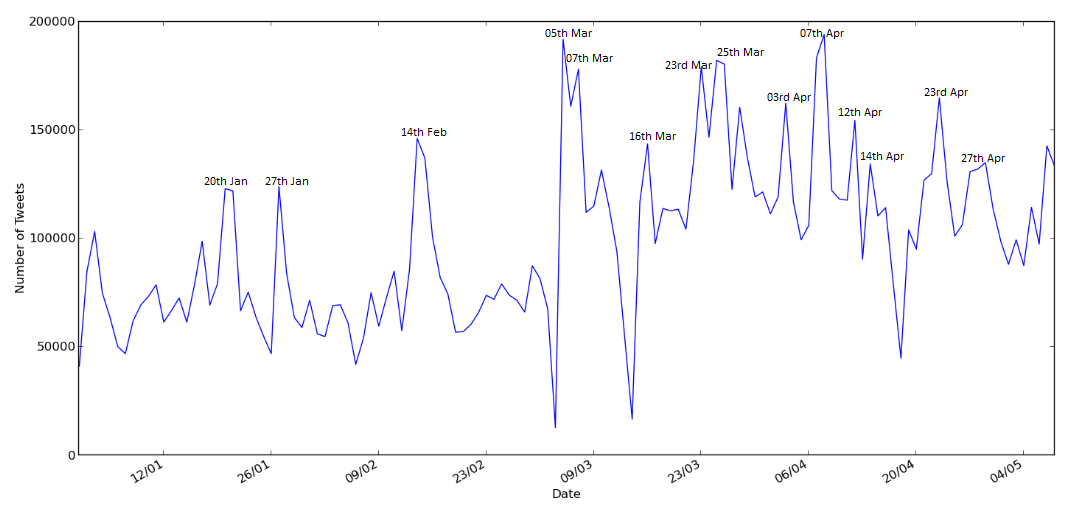}
\caption[Graph depicting the number of tweets per day]{Graph depicting the number of tweets per day}
\label{fig:vol}
\end{figure}

In the Figure \ref{fig:vol}, we find that the number of tweets per day were more in the months of March and April, which is in sync with the claim made above by Gayo-Avello. It would also be important to state that the number of tweets for the months of January and February were about 20 Lac each, whereas the count reached to 36 Lac each for the months of March and April. We next wanted to find out if there were peaks in the volume of the twitter data on days when there was some important activity in the Indian politics regarding the elections. The peaks have been marked with the date in the Figure \ref{fig:vol} and a corresponding table to this timeline is also shown. In the Table \ref{tab:timeline}, we have the event corresponding to the peak and also the most frequent keywords for these dates.

\begin{table}[!h]
\begin{center}
    \begin{tabular}{|l| p {5cm} | p {5cm} |}
    \hline
    \\ \textbf{Date}    & \textbf{Event}	& \textbf{Keywords}	\\ \hline
     20th Jan	      		& Arvind Kejriwal protested at Rail Bhawan, Delhi against the police	        & kejriwal, arvind, bjp, delhi, aap					\\ \hline
    
    27th Jan	           	& Rahul Gandhi's first interview in 10 years on Times Now		    & rahul, gandhi, bjp, \#RahulSpeaksToArnab, interview	 		\\ \hline
    
    14th Feb 				& Arvind Kejriwal resigns as the Chief Minister of Delhi            & kejriwal, bjp, delhi, lokpal, \#aap			 		\\ \hline
    
   05th Mar			   	& Election Commission of India declares Election dates; Arvind Kejriwal visited Gujarat; Rahul Gandhi visited Maharashtra	            & bjp, aap, gandhi, gujarat, \#NaxalAap, \#AKInGujarat, \#AamchaRahul     						\\ \hline
    
    07th Mar				& Last day of Arvind Kejriwal's visit in Gujarat; Kejriwal asked questions to Modi				& bjp, arvind, kejriwal, \#AKasksModi, gujarat \\ \hline
    
    16th Mar				& Modi's declaration as candidate from Varanasi						& bjp, modi, kejriwal, varanasi, contest					\\ \hline
    
    23rd Mar				& Muthalik (accused of assaulting women in pub) joined BJP; ousted next day					& bjp, muthalik, pramod, modi, aap 		\\ \hline
    
    25th Mar				& Kejriwal's rally in Varanasi				& bjp, kejriwal, modi, varanasi, gandhi												\\ \hline
    
    03rd April				& News about rigged EVM in Assam 			& bjp, \#BJPriggedEVM, manifesto, assam, EVM											\\ \hline
    
    07th April				& 1st phase of Elections; BJP released its manifesto				& bjp, manifesto, elections, congress, kejriwal			\\ \hline
    
    12th April				& Rahul Gandhi filed his nomination from Amethi						& gandhi, rahul, modi, \#WeLoveRahul, \#DeshKaNeta				\\ \hline
    
    14th April				& Sonia Gandhi's rally in Aonla				& bjp, gandhi, sonia, kejriwal, pm
   													\\ \hline
   													
   	23rd April				& Priyanka Gandhi's rally in Raibarielly 	& bjp, modi, priyanka, \#PriyanavsBJP, kejriwal								\\ \hline
   	
   	27th April				& A video named `DaamadGate' released by BJP against Robert Vadra	& bjp, gandhi, vadra, priyana, \#DaamadGate					\\ \hline
    
    \end{tabular}
\caption{Timeline of the India Elections stating the important events}\label{tab:timeline}
\end{center}
\end{table}

\begin{figure}[!ht]
\centering
\includegraphics[scale=.5]{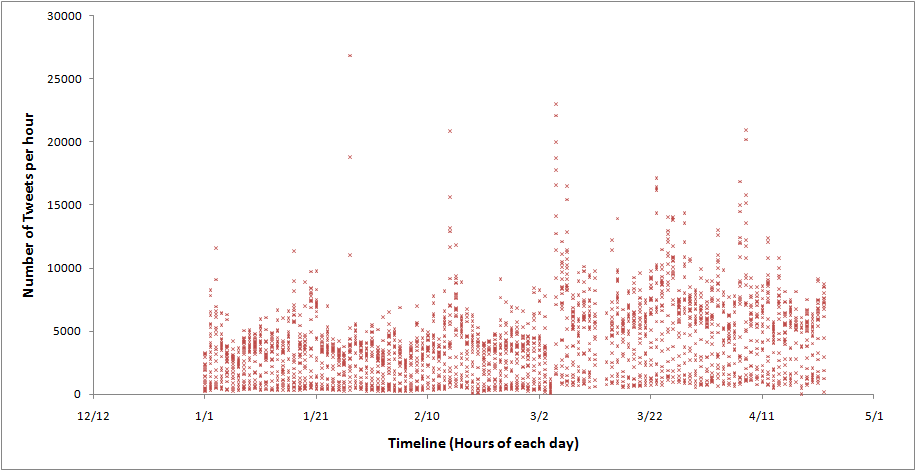}
\caption[Per hour frequency of the number of tweets]{Number of tweets per hour for each day from Jan to Apr}
\label{fig:perhour}
\end{figure}

The next thing that we wanted to see was to see what was the per hour frequency of the tweets, the average, the standard deviation and if the peaks in the per hour frequency of the tweets corresponded to the peaks in the per day frequency graph. To this effect, we plotted the graph given in Figure \ref{fig:perhour} . Here we see that the peaks in the per hour frequency graph correspond exactly to the per day frequency graph. The average number of tweets per hour over this time line is calculated to be 4,073 tweets per hour. Since the highest is more than 27,000 tweets on January 27 and lowest is 1 on April 18, the standard deviation is as high as 3,072 tweets per hour.

\begin{figure}[!ht]
\centering
\includegraphics[scale=.5]{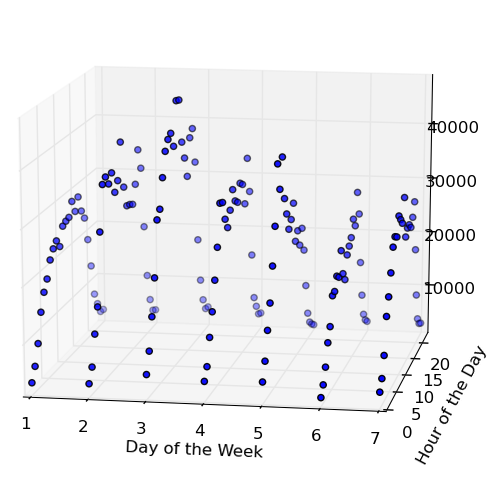}
\caption[Number of tweets for hour of the day v/s day of the week]{Number of tweets for hour of the day v/s day of the week}
\label{fig:dayvsHour}
\end{figure}

We further wanted to see if there were any particular days of week or hours of the day when the tweeting frequency was higher than the usual. For this we developed a 3D graph in Python using the Matplotlib library, see Figure \ref{fig:dayvsHour}. The X axis is the day of the week, the Y axis is the hour of the day and the Z axis is the number of tweets at that particular day of the week and hour of that day. Please note that 1 on the X axis represents Sunday, whereas 7 represents Saturday and the hours on the Y axis are in the 24 hours format.

From this 3D Figure \ref{fig:dayvsHour} we can see that the number of tweets per hour are higher for the second half of the day. The reason for this can be that people are more active during second half of the day as compared to the first half and it peaks during evenings when more people are using Twitter. The election related tweets are particularly higher on Tuesdays and Wednesdays, followed by Mondays and Thursdays. If we look at the days of all the important activities mentioned in the timeline in Table \ref{tab:timeline}, we will see that most of these activities were on weekdays, barring a few exceptions. Also the elections also took place on working days in all the 9 phases, though not all phases are included in this graph. So one can conclude that the Elections related activities were higher during the second half of weekdays, especially Tuesdays and Wednesdays.

\section{User Frequency Analysis}
Since the data we collected was a result of the search query to the streaming API, which returns only about 1\% of the actual data \cite{morstatter2013sample}, we wanted to even see if this 1\% data was good enough. This means that we wanted to see what was the frequency of the tweets made by each user in our dataset. If maximum tweets out of all the tweets were made by only a few users, then this would not have been a good sample. To verify this fact, we found the number of unique users for every month and plotted a graph between the number of users and the number of tweets made by them for each month. Let us look at the  graphs for each month in the following Figure \ref{fig:powerlaw}. We can see that the graphs for each month follows the Power Law and is heavily concentrated at the end of the tail. However, the trail for the months of March and April continue for longer, which means that the number of tweets made by a single user were higher for these months. Few such details have been summarized in the Table \ref{tab:UniqueUsers}.

\begin{figure*}[!h]
  \centering
  \subfigure[]{%
    \includegraphics[width=0.45\textwidth]{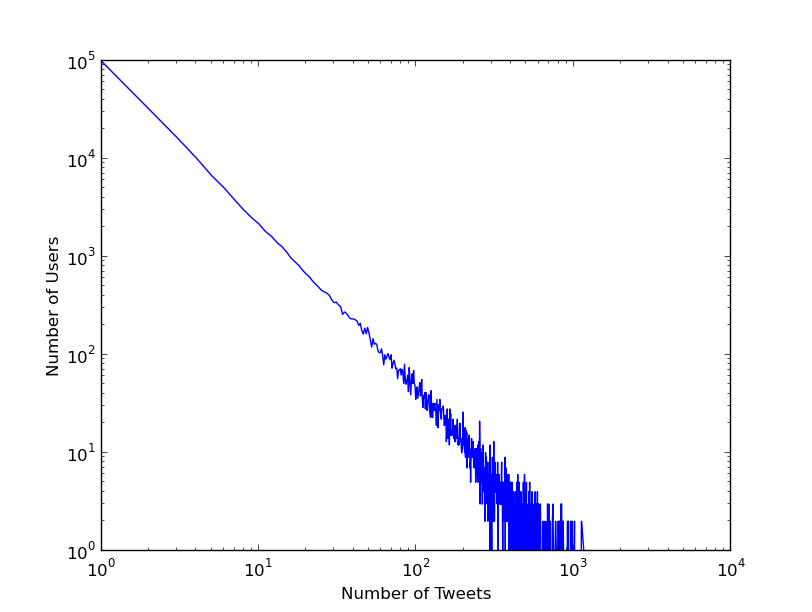}%
    \label{fig:x1}%
  }%
  \hfill
  \subfigure[]{%
    \includegraphics[width=0.45\textwidth]{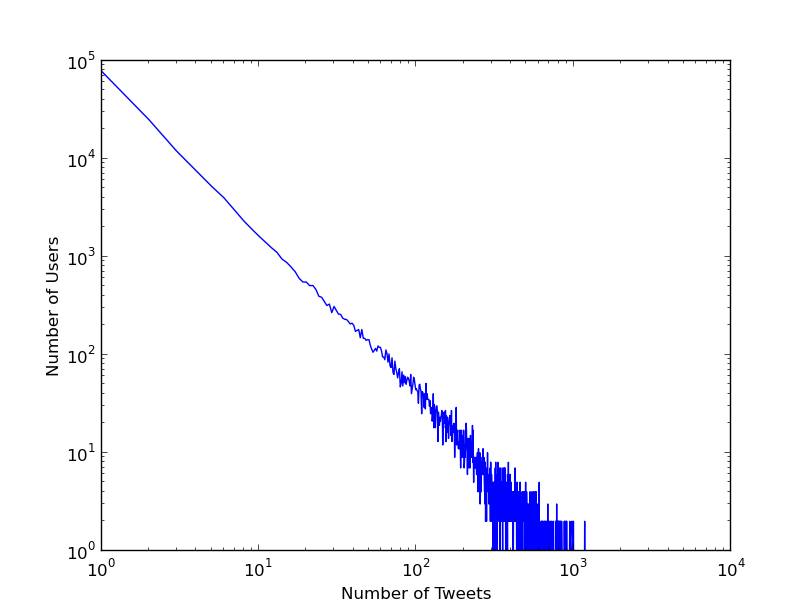}%
    \label{fig:x2}%
  }%
  \hfill
  \subfigure[]{%
    \includegraphics[width=0.45\textwidth]{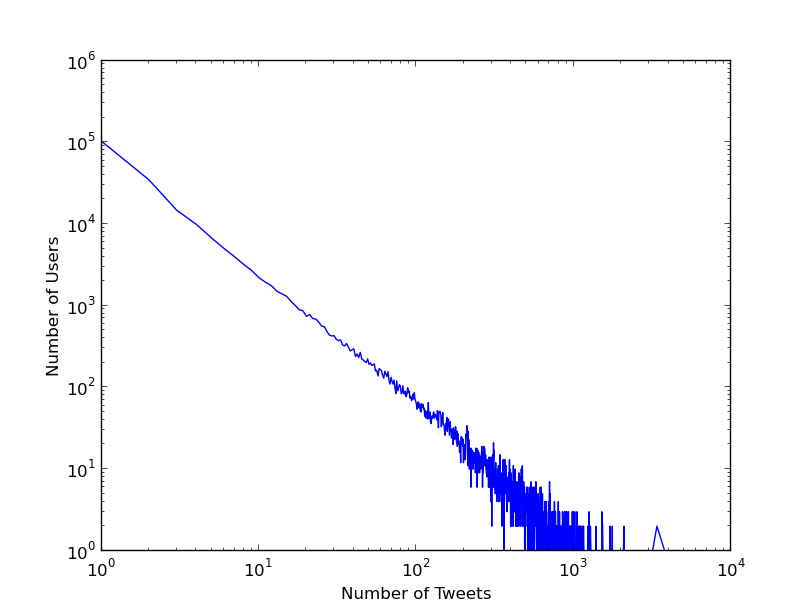}%
    \label{fig:x3}%
  }%
  \hfill
  \subfigure[]{%
    \includegraphics[width=0.45\textwidth]{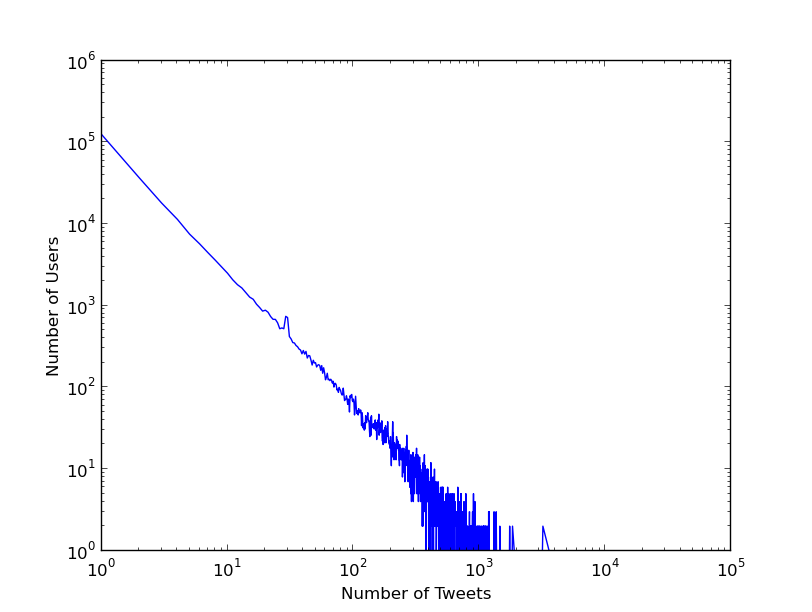}%
    \label{fig:x4}%
  }%
  \caption[Number of users v/s number of tweets graph]{Number of users v/s number of tweets graph for the months of (a) January (b) February (c) March (d) April }
  \label{fig:powerlaw}
\end{figure*}

The data in the Table \ref{tab:UniqueUsers} justifies the long trails for the months of March and April in the adjoining Figure \ref{fig:powerlaw}. The total number of unique users were found to be 815,425 for all these months and \textit{@BJP4India} leading the list with more than 80,000 tweets. Another important observation that we can make from it is that it is nearly the same set of users that have been in the top most active list for all the months, with them exchanging positions among themselves. There is a high probability that these accounts could be bots.

\begin{table}[!h]
\begin{center}
    \begin{tabular}{| >{\centering\arraybackslash}m{1in} | >{\centering\arraybackslash}m{1in} | >{\centering\arraybackslash}m{1in} | >{\centering\arraybackslash}m{2in} |}
    \hline
    \textbf{Month}          & \textbf{\#Tweets in the month}	& \textbf{\#Unique Users}	&\textbf{Top 5 tweeters, \#Tweets}\\ \hline
    \multirow{5}{*}{Jan}   & {\multirow{5}{*}{2,290,785}}	 & {\multirow{5}{*}{209,722}}		  																					& bond\_2014, 6499		\\ 
    					&	&											& rohitagarwal186, 3969 \\
    					&	&											& arshadmohsin, 3172	\\
    					&	&											& dharmava, 2676		\\
    					&	&											& jvsk3, 2569			\\ \hline
    \multirow{5}{*}{Feb}   & {\multirow{5}{*}{2,058,249}}	 & {\multirow{5}{*}{165,189}}		  																					& Tips4DayTrader, 6550	\\ 
    					&	&											& MINDNMONEY, 6414 		\\
    					&	&											& rohitagarwal86, 4466	\\
    					&	&											& Hitarth1987, 3639		\\
    					&	&											& bond\_2014, 3613		\\ \hline
    \multirow{5}{*}{Mar}   & {\multirow{5}{*}{3,653,210}}	 & {\multirow{5}{*}{223,502}}		  																					& bond\_2014, 7336		\\ 
    					&	&											& kavita2680, 6159 		\\
    					&	&											& rohitagarwal86, 6083	\\
    					&	&											& Tips4DayTrader, 6057	\\
    					&	&											& MINDNMONEY, 6042		\\ \hline
    \multirow{5}{*}{Apr}   & {\multirow{5}{*}{3,646,065}}	 & {\multirow{5}{*}{255,906}}		  																					& BJP4India, 34426		\\ 
    					&	&											& rohitagarwal86, 6312 	\\
    					&	&											& bond\_2014, 6061		\\
    					&	&											& Tips4DayTrader, 5889	\\
    					&	&											& MINDNMONEY, 5884		\\ \hline															
    
    \end{tabular}
\caption[Table displaying the number of unique users for each month]{Table displaying the number of tweets, unique users and users with highest number of tweets for each month}\label{tab:UniqueUsers}
\end{center}
\end{table}

\section{Tweet Text Analysis}\label{tweettext}
In this section we shall discuss the next research question. We wanted to see which political party was garnering maximum attention. A very common approach to answer this question is the simple count of the number of mentions. In many papers \cite{nowak1990private}\cite{tumasjan2010predicting}\cite{himelboim2013birds}, this has been used as a method to even predict the elections. So we searched for a set of keywords belonging to each of the parties and on those basis we decided to put a tweet into any of the three buckets. There might be some tweets that fell into more than one categories. 

\begin{figure}[!ht]
\centering
\includegraphics[scale=.6]{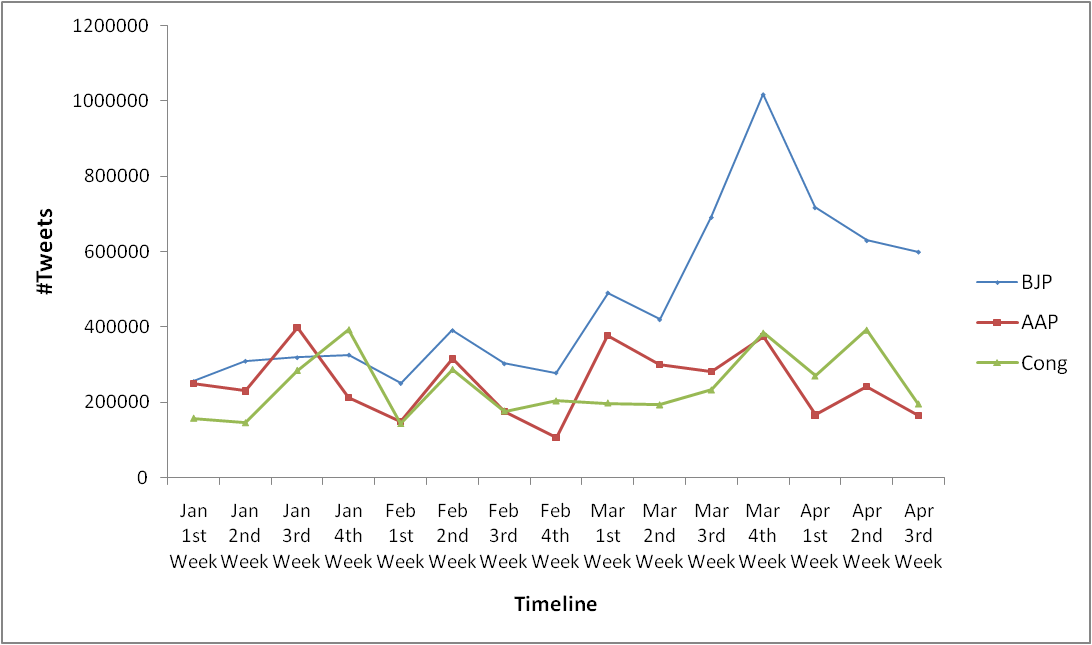}
\caption[Number of tweets related to each party]{Number of tweets related to each party}
\label{fig:mentions}
\end{figure}

\begin{figure}[!ht]
\centering
\includegraphics[scale=.6]{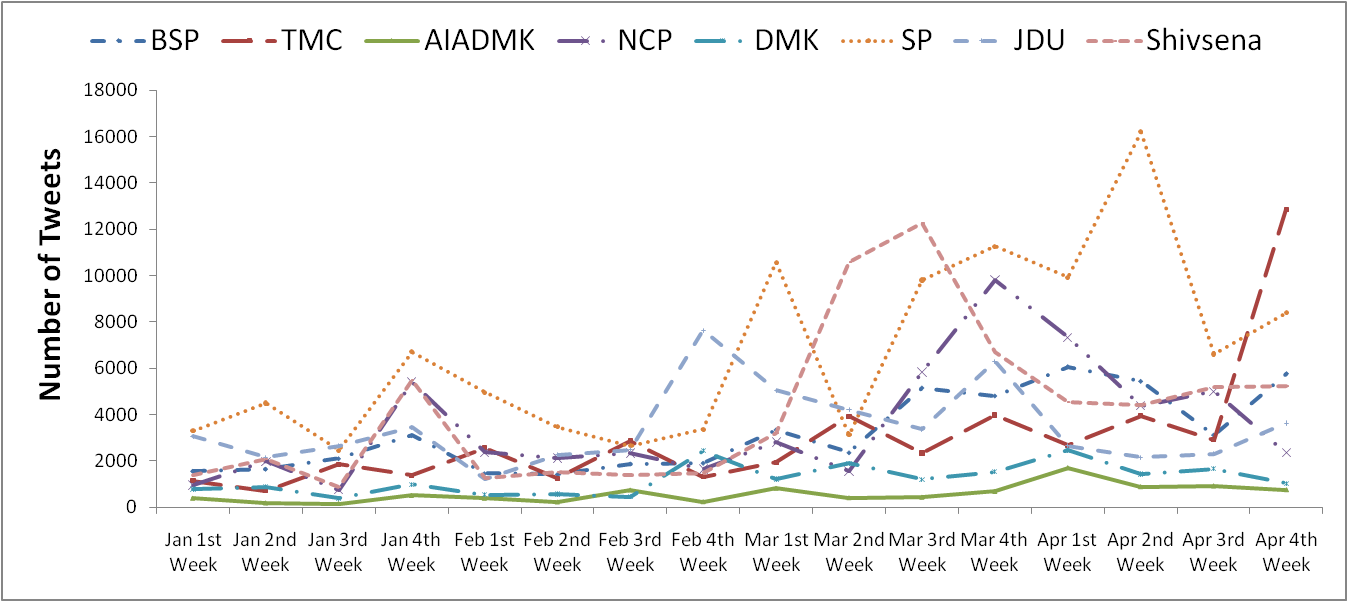}
\caption[Number of tweets related to other party]{Number of tweets related to other party}
\label{fig:otherparties}
\end{figure}

Figure \ref{fig:mentions} shows that during the beginning of the year the number of tweets mentioning about all 3 parties were close enough. The 3rd week of January saw a rise in the number of mentions of AAP when their leader staged a protest in Delhi. Rahul Gandhi's interview with Arnab Goswami made it more popular than any other party in the 4th week of January. Narendra Modi did more than 150 rallies since March, this helped BJP remaining far ahead of any other party for the rest of the weeks. The Congress party started gaining its ground back once the Gandhi family actively started campaigning. Apart from these 3 main parties, we have a graph for the other national and state parties also, shown in Figure \ref{fig:otherparties}. Although, explanation of the timeline of all the parties is well beyond the scope of our work.

\begin{figure}[!ht]
\centering
\includegraphics[scale=.5]{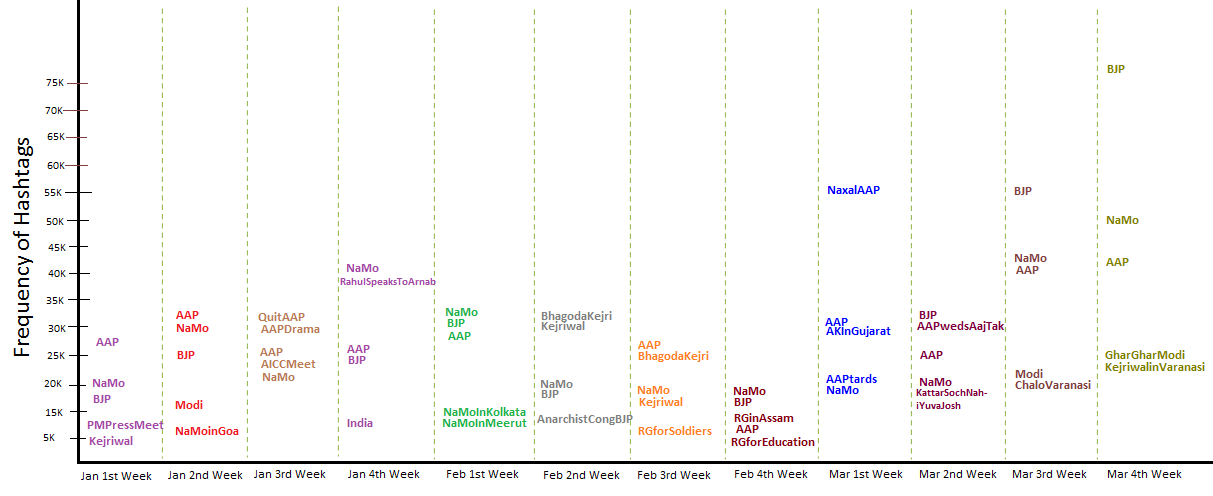}
\caption[Top 5 hashtags for every week]{Top 5 hashtags with their frequencies for each week}
\label{fig:hashtags}
\end{figure}

Since hashtags are an indication of the topic and the mood of the tweet, we wanted to see the top hashtags for each of these weeks. The top 5 hashtags for each week is shown in the Figure \ref{fig:hashtags} along with the frequencies of the hashtag. From this graph we can infer that throughout these 4 months, the major topics have been around the 3 major parties only. Also if we try to annotate these hashtags as Pro or Anti to the major parties, we find that most of the hashtags have been in favor of BJP. The number of Anti and Pro hashtags for AAP have been nearly the same, whereas for Congress, it has been Pro hashtags that seem more like promotion of Rahul Gandhi.

\section{Analyzing the Popularity of Politicians}
In this section we will be comparing the popularity of the two leaders, Arvind Kejriwal (AAP) and Narendra Modi (BJP). We are not taking Rahul Gandhi for the comparison because he does not have a verified account and the account @BeWithRG is also not claimed by him. To compare the popularity, we have taken two measures that are used in calculation of the Klout score as well, i.e. the number of followers and the number of retweets of the tweets made by their account. We calculated the Pearson's correlation factor of each of these two parameters with the Klout score. We found the value of Pearson's correlation between the Klout score and number of followers to be $0.956$, whereas it was $0.463$ only between the Klout score and the average number of retweets. So we can say that the number of followers is a better measure of popularity.

\begin{figure*}[!h]
  \centering
  \subfigure[]{%
    \includegraphics[width=0.7\textwidth]{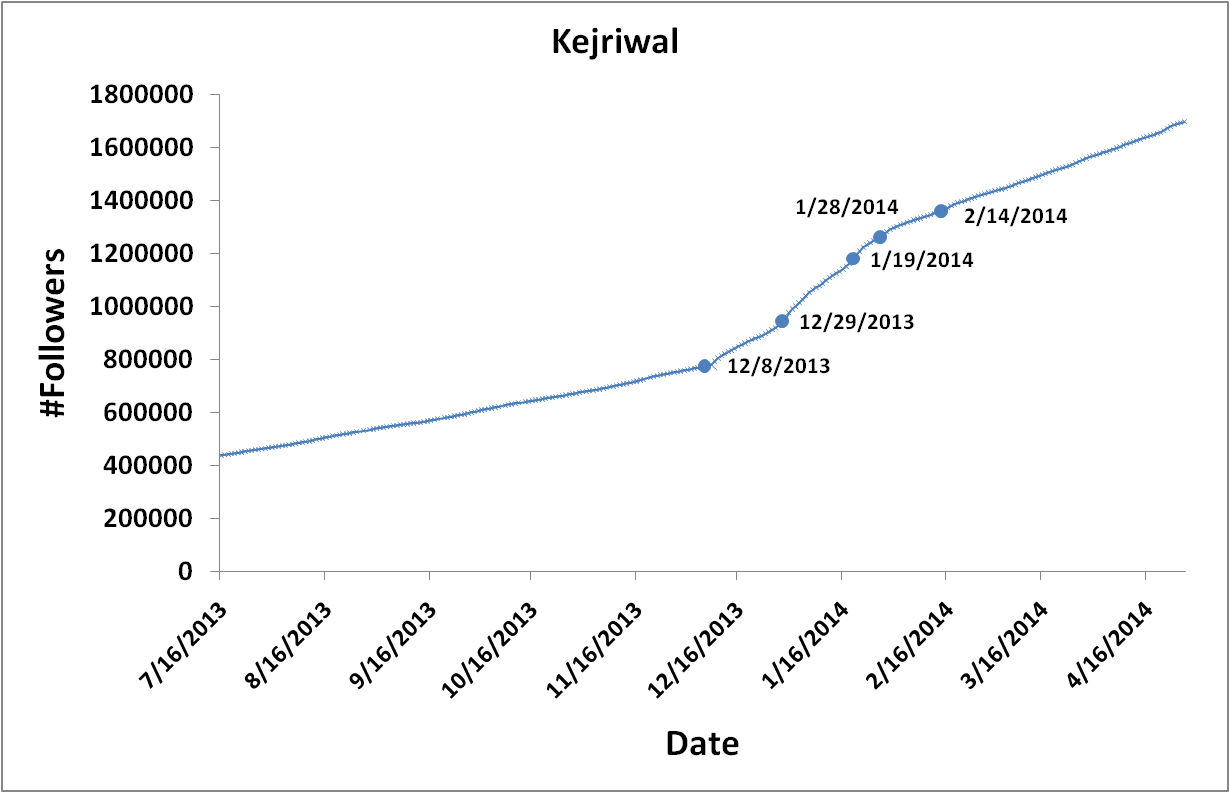}%
    \label{fig:x1}%
  }%
  \hfill
  \subfigure[]{%
    \includegraphics[width=0.7\textwidth]{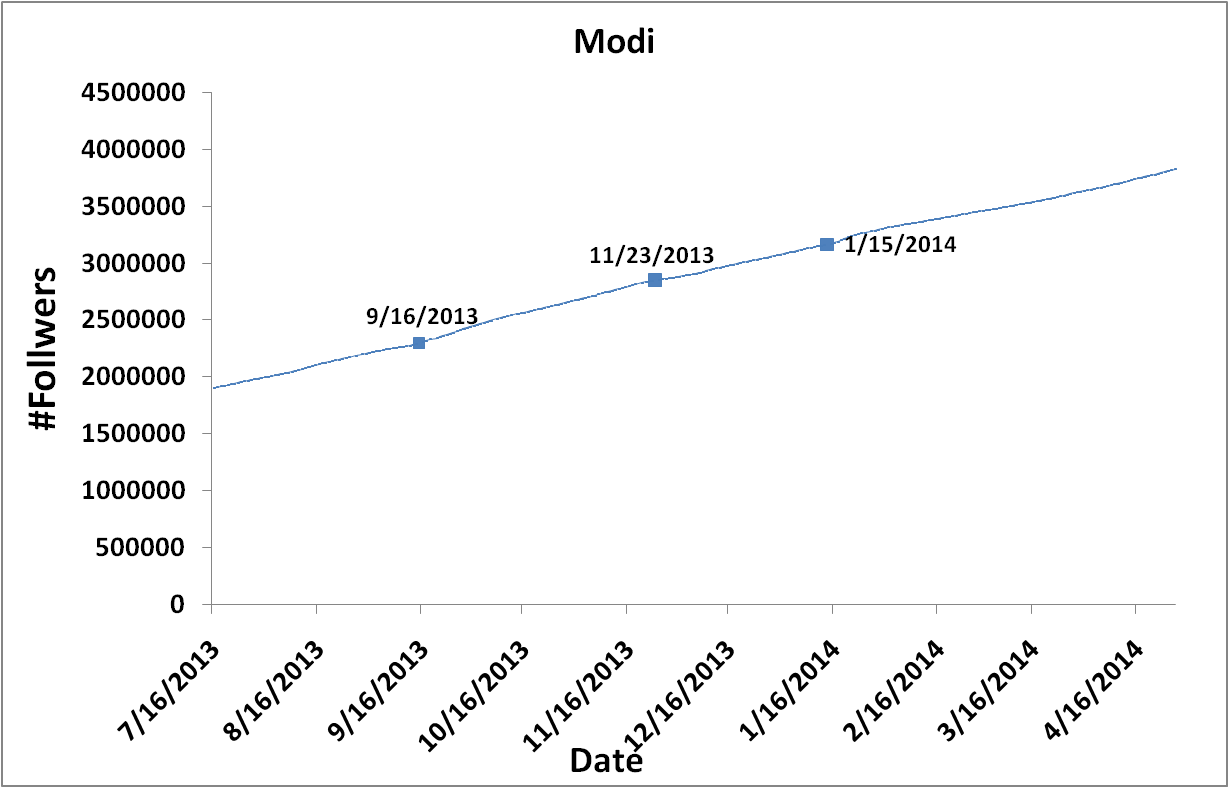}%
    \label{fig:x2}%
  }%
   \caption[Number of followers of Kejriwal and Modi]{Number of followers on daily basis of (a) Kejriwal (b) Modi }
  \label{fig:followercount}
\end{figure*}

We were tracking the accounts of these leaders since long and tried to see their number of followers for each day. The account of Narendra Modi, screen\_name @narendramodi has been on Twitter since January 2009, whereas Arvind Kejriwal's account @ArvindKejriwal has been since November 2011. Graphs showing the number of followers for both these leaders are shown in Figure \ref{fig:followercount}. We can see that the number of followers for Kejriwal in the beginning of the graph is about 0.4 million, whereas it is more than 1.9 million for Modi. There have been steep rise in the graph for Kejriwal around the dates 8th Dec (when he won Delhi assembly elections), 29th Dec (when he took oath as the Delhi CM), 20th Jan (when he staged protest against Delhi police) and the steep rise somewhat normalized after 28th Jan. However there was no fall in the curve around 14th Feb (when he quit as Delhi CM) as was the speculation in the media. When we look at Modi's graph, we do not find any steep rise or fall even after his declaration as pM candidate on 13th Sept, the curve has been constantly increasing except a few glitches on 23rd Nov (which could be due to rise in popularity of AAP).

\begin{figure}[!ht]
\centering
\includegraphics[scale=.6]{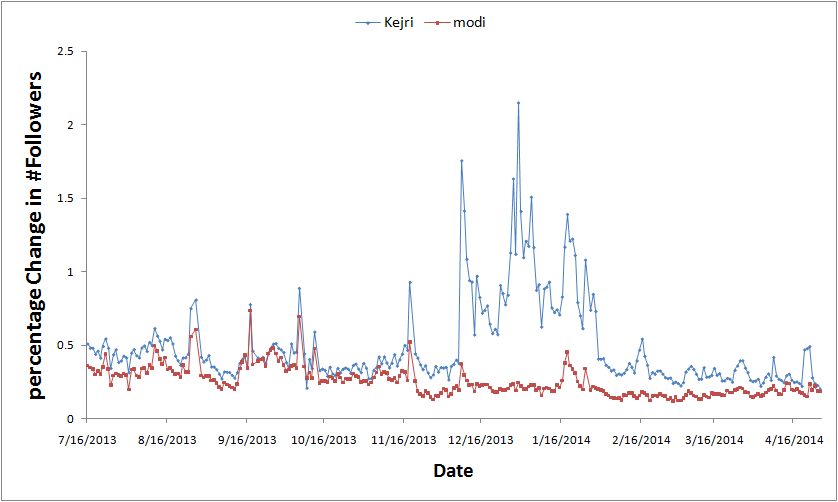}
\caption[Percentage change in number of followers]{Percentage change in number of followers for Kejriwal and Modi}
\label{fig:followerschange}
\end{figure}

To compare the two graphs on a same scale, we compared the percentage change in the number of followers for each day for both these leaders. The graph in Figure \ref{fig:followerschange} shows the comparison. We can see that the change in Kejriwal's followers was always higher than Modi's, which never peaked. Please note that the drop in this graph does not indicate a drop in number of followers, it only shows that the rise in the number of followers was not as high as the previous day. The change was never negative. The average of change in percentage of followers for Kejriwal and Modi were found to be 0.49\% and 0.25\% respectively. The average number of followers per day for Kejriwal was 1,372, whereas it was 2,020 for Modi. So if we look at it with this aspect, Modi has more followers and thus more popular.

\begin{figure*}[!ht]
  \centering
  \subfigure[]{%
    \includegraphics[width=0.45\textwidth]{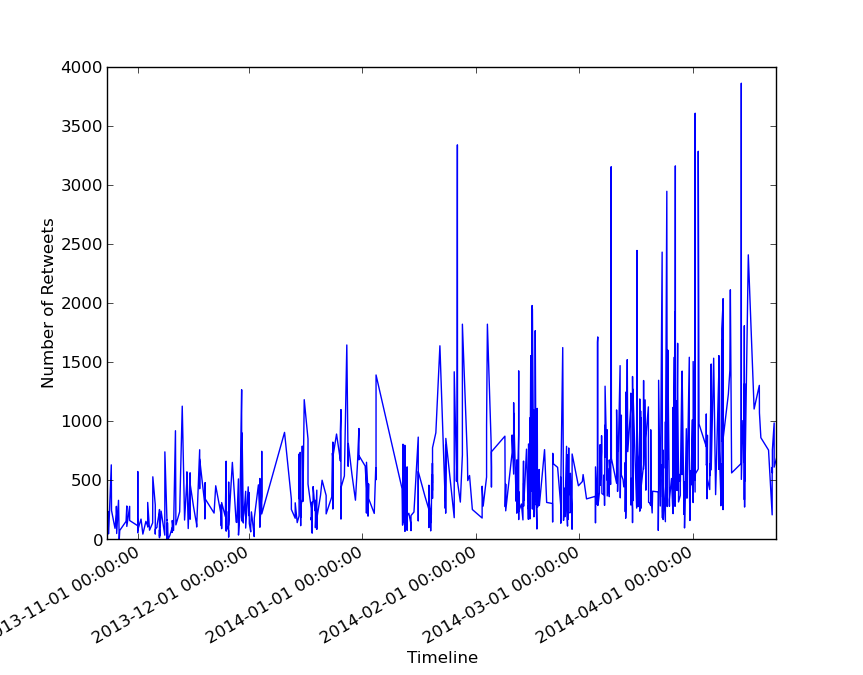}%
    \label{fig:kejriretweet}%
  }%
  \hfill
  \subfigure[]{%
    \includegraphics[width=0.45\textwidth]{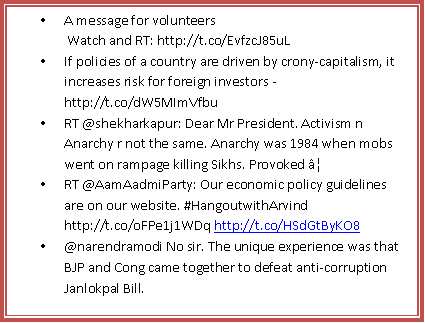}%
    \label{fig:kejriretweettext}%
  }%
  \hfill

  \subfigure[]{%
    \includegraphics[width=0.45\textwidth]{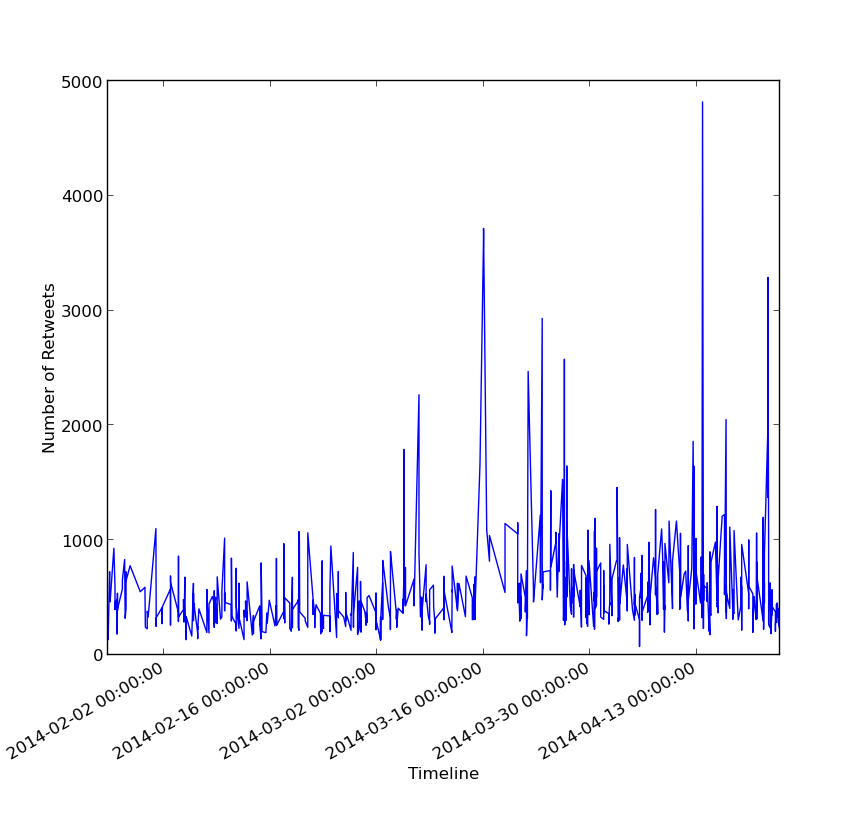}%
    \label{fig:modiretweet}%
  }%
  \hfill
  \subfigure[]{%
    \includegraphics[width=0.45\textwidth]{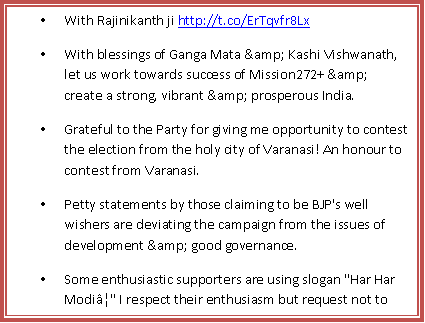}%
    \label{fig:modiretweettext}%
  }%
   \caption[Retweet count of Kejriwal and Modi]{(a) Retweet count on tweets Kejriwal (b) Kejriwal's tweets which received maximum retweets (c) Retweet count on tweets Modi (d) Modi's tweets which received maximum retweets }
  \label{fig:retweet}
\end{figure*}

We would now look at the retweet count of the tweets made by Kejriwal and Modi. For this we took last 1,000 tweets made by both these leaders and retrieved the retweet\_count from the JSON object and plotted it on a graph against the timeline. This graph can be found in the Figure \ref{fig:retweet}. Though Modi has the maximum retweet count of 5,000 on a tweet, the retweet count for Kejriwal's retweets are generally higher. This can also be established by comparing the average retweet count for their last 1,000 tweets, which is 887 for Arvind Kejriwal, whereas 612 for Modi. So with this aspect, Kejriwal is more popular than Modi. But since we claimed that the number of followers is a rather higher correlated parameter to the Klout score, we can conclude that Modi had better popularity than Kejriwal.

Since we were tracking the accounts of various parties and politicians, one thing that caught our attention was the number of tweets made by the account of BJP, named @BJP4India. The tweet count of this account went from 2538 on Feb 01 to 96,462 on Apr 30. They made more than 20,000 on Apr 07 and more than 100\% rise in tweets on several occasions. This clearly shows the way BJP was proactively promoting itself and its PM candidate on twitter, which could be a possible reason for their lead in almost all graphs.

We also tried to look at the possibility of the accounts being \textbf{bot accounts}. For this purpose, we used a tool by Indiana University.\footnote{\url{http://truthy.indiana.edu/botornot/}} This tool took the screen\_names of the accounts as input and the analyzed the account for being a bot or not. The conclusion was made on the basis of network analysis, the tweets made and some other factors. We used this tool to analyze some 50 odd profiles from our dataset. And we found out that almost all the profiles were not bots. The possible reason for this could be that since we had hand picked these profiles, the chances of them being bots was less. Apart from this we also checked for a few profiles which were tweeting about elections and found that profiles like Tips4DayTrader were found to be bots. 

\chapter{Identifying Political Orientation: Experiment and Analysis} \label{chp6}
Several papers have tried to predict the political alignment of users on Twitter. Some attempts such as Conover  et. al \cite{conover2011predicting} achieved an efficiency of 95\%, highest till date, other attempts such as Cohen et. al could achieve barely 65\% efficiency. This difference in efficiency in different papers inspired us to check what was the efficiency for the users involved in India Elections 2014. As explained in the Section \ref{methodology}, we collected the information about 1000 random users from our data set who were tweeting about elections and got them annotated with a help of data annotators. This gave us a true positive data set and we then used the following four techniques on this data to develop an efficient classifier. The results of each techniques have been mentioned along with them.

\section{Content Based Techniques}
In this Section, we shall be discussing the two techniques that were based on the content generated by the users in the form of tweets. A tweet is a 140-character message that is composed and posted by the user on Twitter.\footnote{\url{https://support.twitter.com/articles/15367-posting-a-tweet}} This tweet may or may not contain the personal views of the user. But since the orientation of the users were decided by the annotators on the basis of the tweets made by them, it was imperative to develop a classifier based on this content. The tweet may have several parts: text, URLs, hashtags, user mentions etc. The two methods discussed in the following subsections make use of these parts of the content of the tweet.

\subsection{Text Based Classification}
In this method, we made use of the words in the tweets. We took 200 tweets from the user's timeline between March 20 and April 10. This was the time when our annotators were annotating and deciding the orientation of the users. So the tweets that were available in our dataset and the ones on the basis of which our annotators decided the orientation were nearly the same. 

In this technique, we removed the hashtags, URLs, user mentions and other common words such as `RT', `\&amp', `\textit{ka}', `\textit{ke}', `\textit{ki}' etc. from the tweet text. After removing these words, we made a user matrix of the TF-IDF of the words used in the tweets by the users. Let us now understand the calculation of this vector. To understand this, we need to understand TF-IDF first. TF-IDF, abbreviation for Term Frequency- Inverse Document Frequency is a measure of the importance of a term in a document in a collection of documents or corpus \cite{wu2008interpreting}. In our case, the term frequency is the ratio of the number of times a word or a term was used in the 200 tweets of the user to the total number of terms produced by the user in those 200 tweets. Therefore,
\begin{equation}\label{equation:6.1}
TF_{\textit{i,j}}= \frac{n_\textit{i,j}}{\sum_{k} n_{\textit{k,j}}}
\end{equation}

The Equation \ref{equation:6.1} has been picked from Conover et. al \cite{conover2011predicting}. For the \textit{i$_{th}$} term of the user \textit{j}, n$_{\textit{i,j}}$ is the number of times the term \textit{i} occurs in the tweets of user \textit{j}. And $\sum_{k} n_{k,j}$ is the sum of total terms produced by the user \textit{j} over \textit{k} tweets. Since IDF is the measure of how common is the word in the corpus, IDF in our case is the logarithmic ratio of the total number of users by the number of users who produce term \textit{i}. 

\begin{equation} \label{eq:eq62}
IDF_{i} = \log{\frac{|{U}|}{1+|U_{i}|}}
\end{equation}
The TFIDF is calculated as the product of the TF and IDF, i.e.,

\begin{equation}
TFIDF_{i,j} =  TF_{i,j} \bullet IDF_{i}
\end{equation}

While we were calculating the IDF for various terms, we removed those terms which were not used by atleast 5 users. Such words do not contribute to the efficiency of the performance of the classifier, but only increase memory usage. Please note that we added 1 in the denominator of the Equation \ref{eq:eq62} to bring the difference between the words that were produced by all the users and the words which were not produced by the user \textit{i}. We fed the true positive data we had to this algorithm to generate the TFIDF vector. We initially used all the 613 users from the `Pro' category we had. These 613 users were the ones whose `Pro' political orientation was known to us and not `CAN'T SAY'. All the experiments in this chapter have been conducted with the help of Weka 3.6.6. The classifier used in most of the cases has been Random Forest with 10 fold cross-validation. Weka uses the default split of 66\% of training data and 33\% of testing data for each fold. Though we experimented with other classifiers as well, but have reported the results for this one classifier only to maintain uniformity. Results for this experiment have been specified in the Table \ref{tab:result1}.

\begin{table}[!h]
\begin{center}
    \begin{tabular}{|l|l|l|l|}
    \hline
    \multicolumn{4}{|c|}{\textbf{Instances:} 613}											\\ \hline
    \multicolumn{4}{|c|}{\textbf{Attributes:} 9312}											\\ \hline
    \multicolumn{4}{|c|}{\textbf{Classifier:} Random Forest, 10 folds cross-validation}		\\ \hline
    \multicolumn{4}{|c|}{\textbf{Efficiency:} 72.36\%}										\\ \hline
    \textbf{Party}          & \textbf{Precision}	& \textbf{Recall}	& \textbf{F-measure}\\ \hline
    \textbf{AAP}      		& 0.381					& 0.061				& 0.105 	     	\\ \hline
    \textbf{BJP}           	& 0.736					& 0.975				& 0.839      		\\ \hline
    \textbf{CONG} 			& 0					    & 0    				& 0     			\\ \hline
    \end{tabular}
\caption[Experiment results with all 613 users of `Pro' category]{Experiment results with all 613 users of `Pro' category}\label{tab:result1}
\end{center}
\end{table}

The efficiency obtained with all 613 instances was good enough, but the precision and recall values for AAP and Congress were not good. We assumed that this could be because of the unbalanced data that was fed into the classifier. To overcome this problem, we experimented with equal number of instances for all classes. Since we had 33 users `Pro' to Congress, we took 33 instances of AAP and BJP and did the same experiment as above. The results have been summarized in Table \ref{tab:result2}. Here we can see that although the precision and recall values for both AAP and Congress have improved, but the overall efficiency reduced significantly and below acceptable limits.

\begin{table}[!h]
\begin{center}
    \begin{tabular}{|l|l|l|l|}
    \hline
    \multicolumn{4}{|c|}{\textbf{Instances:} 99}											\\ \hline
    \multicolumn{4}{|c|}{\textbf{Attributes:} 2442}											\\ \hline
    \multicolumn{4}{|c|}{\textbf{Classifier:} Random Forest, 10 folds cross-validation}		\\ \hline
    \multicolumn{4}{|c|}{\textbf{Efficiency:} 42.42\%}										\\ \hline
    \textbf{Party}          & \textbf{Precision}	& \textbf{Recall}	& \textbf{F-measure}\\ \hline
    \textbf{AAP}      		& 0.424					& 0.758				& 0.543 	     	\\ \hline
    \textbf{BJP}           	& 0.481					& 0.394				& 0.433      		\\ \hline
    \textbf{CONG} 			& 0.308				    & 0.121				& 0.174    			\\ \hline
    \end{tabular}
\caption[Experiment results with equal instances of all classes in `Pro' category]{Experiment results with equal instances of all classes in `Pro' category}\label{tab:result2}
\end{center}
\end{table}
We conducted the same experiment with different combinations of instances, but never touched a mark of 70\% efficiency. The precision and recall values were also low for some or the other class every time. We then did this experiment for only 2 classes to see if the results were any better. The highest efficiency that we could achieve was with 33 instances of AAP and 33 instances of Congress- 65.15\%. With 130 instances each of BJP and AAP, we got an efficiency of 59.31\%. The combination of BJP and Congress however gave us a low efficiency score of 46.96\% and precision and recall values less than 0.5. We then moved on to conducting the same experiment for the `Anti' category users. We first did it with all 425 instances and then with 85 instances of each class. The two results have been summarized in Tables \ref{tab:result3} and \ref{tab:result4}.

\begin{minipage}[b]{.45\linewidth}
  \centering
    \begin{tabular}{|m{1cm}|m{1cm}|m{1cm}|m{1.5cm}|}
    \hline
    \multicolumn{4}{|c|}{\textbf{Instances:} 425}											\\ \hline
    \multicolumn{4}{|c|}{\textbf{Attributes:} 8014}											\\ \hline
    \multicolumn{4}{|c|}{\textbf{Classifier:} Random Forest}								\\ \hline
    \multicolumn{4}{|c|}{\textbf{Efficiency:} 47.75\%}										\\ \hline
	\textbf{Party}  & \textbf{Pre}	& \textbf{Recall}	& \textbf{F-measure}				\\ \hline
    AAP      		& 0.489					& 0.863				& 0.624 	     			\\ \hline
    BJP           	& 0.313					& 0.059 			& 0.099      				\\ \hline
    CONG 			& 0.447				    & 0.157				& 0.232    					\\ \hline    
    \end{tabular}
\captionof{table}{Classification results with all instances of `Anti' category}\label{tab:result3}
\end{minipage}
\hspace{5mm}
\begin{minipage}[b]{.45\linewidth}
  \centering
    \begin{tabular}{|m{1cm}|m{1cm}|m{1cm}| m{1.5cm}|}
    \hline
    \multicolumn{4}{|c|}{\textbf{Instances:} 255}											\\ \hline
    \multicolumn{4}{|c|}{\textbf{Attributes:} 6847}											\\ \hline
    \multicolumn{4}{|c|}{\textbf{Classifier:} Random Forest}								\\ \hline
    \multicolumn{4}{|c|}{\textbf{Efficiency:} 37.25\%}										\\ \hline
    \textbf{Party}  & \textbf{Pre}			& \textbf{Recall}	& \textbf{F-measure}		\\ \hline
    AAP      		& 0.321					& 0.529				& 0.4 	     				\\ \hline
    BJP           	& 0.47					& 0.365				& 0.411      				\\ \hline
    CONG 			& 0.388				    & 0.224				& 0.284    					\\ \hline
    \end{tabular}
\captionof{table}[Experiment results with equal instances of all classes in `Anti' category]{Experiment results with equal instances of all classes in `Anti' category}\label{tab:result4}
\end{minipage}

From all these results, we can conclude that a two class classification based on the text of the tweets can be easy for AAP \& Congress and AAP \& BJP, but when we try a 3-class classification with all three parties, the results are not favorable.

\subsection{Hashtags Based Classification}
Since the results with text based classification were not satisfactory, we tried another methodology. This time, we used the hashtags present in the tweets. We extracted the hashtags from the last 200 tweets of each user and calculated the term frequency of each the hashtag used by a user \textit{j}. Next the IDF of the hastag \textit{i} was calculated in exactly the similar manner as above. The purpose of using this method was the assumption that since hashtags are used to represent the topic of the tweet, the noise in the data would be reduced. Another important thing to be noted here is that, in this method we removed those hashtags which were used by only 1 user, in comparison to the previous method where a word was included in the vector only if it was used by atleast 5 people. The explanation  for this change is that the number of hashtags used in a tweet are drastically lower than the total number of words used in the tweets.

As in the previous case, we first tried the classifier by feeding all the instances of `Pro' category that we had. This gave us 1398 attributes with 613 instances of all 3 parties. The results have been summarized in Table \ref{tab:result5}. If we compare the efficiency of this method with that of text based classification in Table \ref{tab:result1}, we see that this method gives us a higher efficiency and is in line with our thoughts. However, the precision and recall values in this case also are not very impressive for AAP and Congress.

\begin{minipage}[t]{.45\linewidth}
  \centering
    \begin{tabular}{|m{1cm}|m{1cm}|m{1cm}|m{1.5cm}|}
    \hline
    \multicolumn{4}{|c|}{\textbf{Instances:} 613}											\\ \hline
    \multicolumn{4}{|c|}{\textbf{Attributes:} 1398}											\\ \hline
    \multicolumn{4}{|c|}{\textbf{Classifier:} Random Forest}								\\ \hline
    \multicolumn{4}{|c|}{\textbf{Efficiency:} 75.49\%}										\\ \hline
	\textbf{Party}  & \textbf{Pre}	& \textbf{Recall}	& \textbf{F-measure}				\\ \hline
    AAP      		& 0.759					& 0.167				& 0.273 	     			\\ \hline
    BJP           	& 0.756					& 0.983 			& 0.856      				\\ \hline
    CONG 			& 0					    & 0					& 0	    					\\ \hline    
    \end{tabular}
\captionof{table}{Hashtag based classification results with all instances of `Pro' category}\label{tab:result5}
\end{minipage}
\hspace{5mm}
\begin{minipage}[t]{.45\linewidth}
  \centering
    \begin{tabular}{|m{1cm}|m{1cm}|m{1cm}| m{1.5cm}|}
    \hline
    \multicolumn{4}{|c|}{\textbf{Instances:} 99}											\\ \hline
    \multicolumn{4}{|c|}{\textbf{Attributes:} 252}											\\ \hline
    \multicolumn{4}{|c|}{\textbf{Classifier:} Random Forest}								\\ \hline
    \multicolumn{4}{|c|}{\textbf{Efficiency:} 43.43\%}										\\ \hline
    \textbf{Party}  & \textbf{Pre}			& \textbf{Recall}	& \textbf{F-measure}		\\ \hline
    AAP      		& 0.391					& 0.818				& 0.529 	     			\\ \hline
    BJP           	& 0.476					& 0.303				& 0.37      				\\ \hline
    CONG 			& 0.667				    & 0.182				& 0.286    					\\ \hline
    \end{tabular}
\captionof{table}{Hashtag based classification results with equal instances of all classes in `Pro' category}\label{tab:result6}
\end{minipage}

In this method also we tried 2-class classification for different pairs and the results again showed that the AAP-BJP efficiency was the highest with 60\%, followed by the AAP-Congress pair with 59.09\% efficiency and a close third was the BJP-Congress pair with 57.57\%. Precision and recall values in all the cases was close to 0.5, thus proving this method to be better than the text based classification method.

For the `Anti' category of users also, we found that the efficiency was higher when compared with the text based classification results. The results for all 425 instances as well as 255 instances (with 85 for each class) have been shown in Tables \ref{tab:result7} and \ref{tab:result8}.

\begin{minipage}[t]{.45\linewidth}
  \centering
    \begin{tabular}{|m{1cm}|m{1cm}|m{1cm}|m{1.5cm}|}
    \hline
    \multicolumn{4}{|c|}{\textbf{Instances:} 425}											\\ \hline
    \multicolumn{4}{|c|}{\textbf{Attributes:} 1182}											\\ \hline
    \multicolumn{4}{|c|}{\textbf{Classifier:} Random Forest}								\\ \hline
    \multicolumn{4}{|c|}{\textbf{Efficiency:} 50.35\%}										\\ \hline
	\textbf{Party}  & \textbf{Pre}	& \textbf{Recall}	& \textbf{F-measure}				\\ \hline
    AAP      		& 0.5					& 0.946				& 0.654 	     			\\ \hline
    BJP           	& 0.875					& 0.165 			& 0.277      				\\ \hline
    CONG 			& 0.286				    & 0.045				& 0.077	    					\\ \hline    
    \end{tabular}
\captionof{table}{Hashtag based classification results with all instances of `Anti' category}\label{tab:result7}
\end{minipage}
\hspace{5mm}
\begin{minipage}[t]{.45\linewidth}
  \centering
    \begin{tabular}{|m{1cm}|m{1cm}|m{1cm}| m{1.5cm}|}
    \hline
    \multicolumn{4}{|c|}{\textbf{Instances:} 255}											\\ \hline
    \multicolumn{4}{|c|}{\textbf{Attributes:} 716}											\\ \hline
    \multicolumn{4}{|c|}{\textbf{Classifier:} Random Forest}								\\ \hline
    \multicolumn{4}{|c|}{\textbf{Efficiency:} 34.25\%}										\\ \hline
    \textbf{Party}  & \textbf{Pre}			& \textbf{Recall}	& \textbf{F-measure}		\\ \hline
    AAP      		& 0.325					& 0.318				& 0.321 	     			\\ \hline
    BJP           	& 0.412					& 0.329				& 0.366      				\\ \hline
    CONG 			& 0.311				    & 0.381				& 0.342    					\\ \hline
    \end{tabular}
\captionof{table}{Hashtag based classification results with equal instances of all classes in `Anti' category}\label{tab:result8}
\end{minipage}

For the testing purpose, we did the 2-class classification as well and got the efficiency to be 52.94\% for the AAP-BJP pair with 85 instances of each class. Please note, that we tried all the above stated methods with different combinations of instances of each class, viz., taking the data in 1:2:2 (CONGRESS:BJP:AAP) or 1:2:3; but none resulted in a good efficiency coupled with high precision and recall values. Hashtag based classification for the `Anti' category though has some improvement than the text based classification, however an unbiased sample of data gives a very low efficiency and thus it can be claimed that in the Indian Elections, classification of users tweeting about elections on the basis of hashtags would not give us a good proficiency. 

\section{User Features Based Classification}
After having very low precision and recall values and moderate accuracy in both the methods of the content based techniques, we decided to move to a different methodology. This time we decided to use some of the user features as the features for training and testing our classifier. The features we included were:

\begin{itemize}
\item Number of friends
\item Number of followers
\item Following AAP?
\item Following BJP?
\item Following Congress?
\item Number of AAP related words
\item Number of BJP related words
\item Number of Congress related words
\item Number of AAP related hashtags
\item Number of BJP related hashtags
\item Number of Congress related hashtags
\end{itemize}

\begin{figure}[!ht]
\centering
\includegraphics[scale=.6]{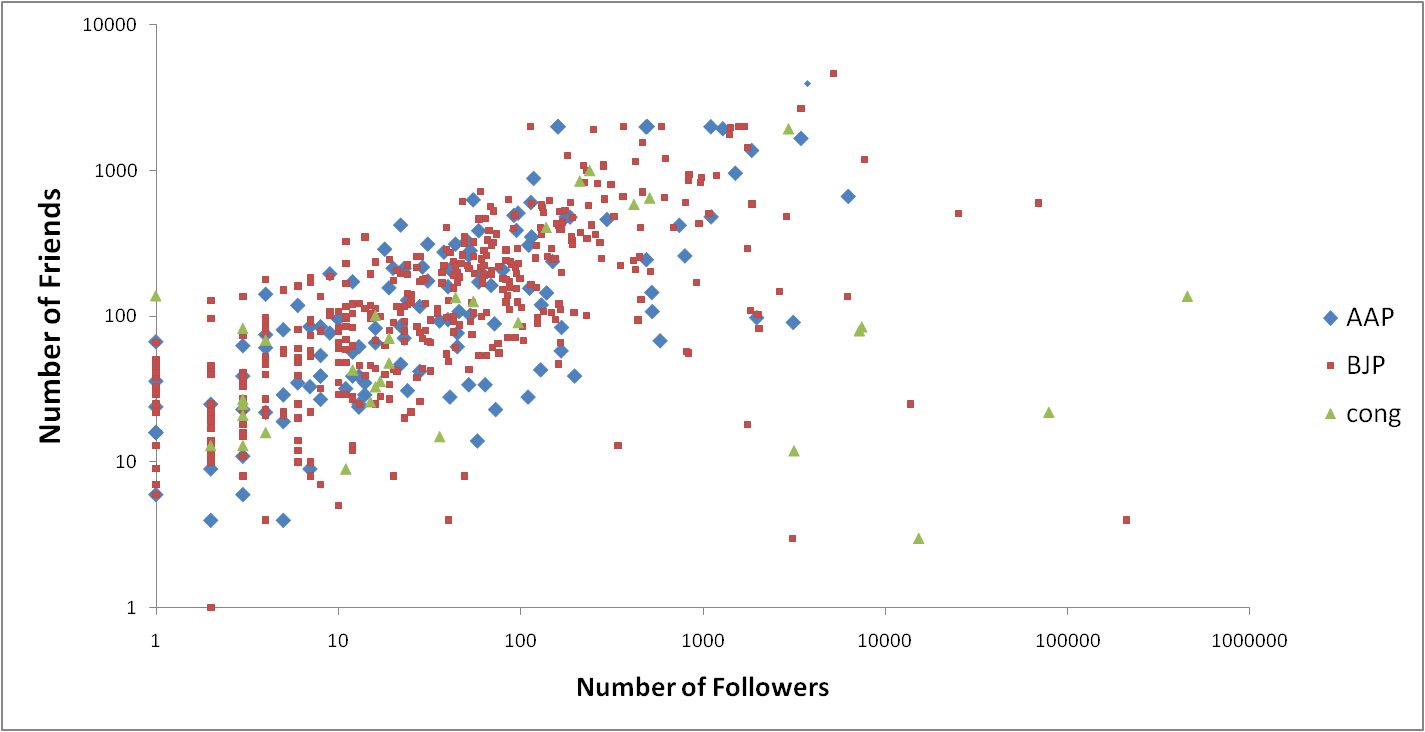}
\caption[Scatter plot of followers v/s friends count]{Scatter plot of followers v/s friends count for all 3 parties}
\label{fig:scatter}
\end{figure}

Let us understand the intuition behind including these features and not others. The first two features are the number of friends and the number of followers. This feature is regarding the number of other twitter profiles that the user follows and is being followed by. We plotted a scatter plot for friends v/s followers for different classes and tried to observe the difference shown in Figure \ref{fig:scatter}. The diamond shaped data points represent the number of friends and followers for those users who were `Pro' to AAP, the square shows it for BJP and triangle for Congress. We can see that most of the outliers belong to BJP and Congress. One can see that most of the data points that belong to BJP and Congress have a very high followers count as compared to friends count despite the fact that not many of these data points represent some popular politicians who might be having a high followers count and low number of friends. The average number of followers and friends for AAP, BJP and Congress were 264, 283, 911, 233, 17,216, 210 respectively. This clearly shows the anomalies which could be used as a strong feature for classifying the data.

The next set of features that we took were who was the user following. We wanted to see if the user was following AAP, BJP or Congress. Since one can be following more than 1, and that would have made it a multi-valued attribute. To avoid this, we made 3 features out of it: following AAP, following BJP and following Congress. Each of these 3 attributes was a binary valued attribute. The value of these attributes was 1 if the user was following any of the accounts related to AAP, BJP or Congress and 0 otherwise. There were buckets of accounts related to each of these 3 parties. If a user was following even one of the accounts from that bucket, the value of the attribute related to that party was set to 1. The accounts in these buckets were chosen from the list of politicians we were tracking (see Section \ref{trending politicians}). Since all these accounts were handpicked, we segregated them to be belonging to these 3 parties and we then extracted their IDs. The next 3 features were the number of times the users mentioned the words related to AAP, BJP and Congress. And the last 3 features were the number of times the user made use of hashtags related to the 3 parties. These hashtags were chosen on the basis of top 10 hashtags that we discussed in Section \ref{tweettext}. The reason for including these last 6 features was to include the effect of the content based techniques as well. 

As in the previous methodologies, we fed the classifier with all 613 instances of `Pro' class to look at the efficiency. The results have been summarized in Table \ref{tab:result9}. One interesting thing about the results is that even though the efficiency is bit lower than the hashtag based classification, but the precision and recall values of Congress in this case are not 0. This makes this method better than the hashtag based classification.

\begin{minipage}[t]{.45\linewidth}
  \centering
    \begin{tabular}{|m{1cm}|m{1cm}|m{1cm}|m{1.5cm}|}
    \hline
    \multicolumn{4}{|c|}{\textbf{Instances:} 613}											\\ \hline
    \multicolumn{4}{|c|}{\textbf{Attributes:} 12}											\\ \hline
    \multicolumn{4}{|c|}{\textbf{Classifier:} Random Forest}								\\ \hline
    \multicolumn{4}{|c|}{\textbf{Efficiency:} 71.45\%}										\\ \hline
	\textbf{Party}  & \textbf{Pre}	& \textbf{Recall}	& \textbf{F-measure}				\\ \hline
    AAP      		& 0.455					& 0.376				& 0.412 	     			\\ \hline
    BJP           	& 0.781					& 0.855 			& 0.816      				\\ \hline
    CONG 			& 0.429				    & 0.182				& 0.255    					\\ \hline    
    \end{tabular}
\captionof{table}{User features based classification results with all instances of `Pro' category}\label{tab:result9}
\end{minipage}
\hspace{5mm}
\begin{minipage}[t]{.45\linewidth}
  \centering
    \begin{tabular}{|m{1cm}|m{1cm}|m{1cm}| m{1.5cm}|}
    \hline
    \multicolumn{4}{|c|}{\textbf{Instances:} 99}											\\ \hline
    \multicolumn{4}{|c|}{\textbf{Attributes:} 12}											\\ \hline
    \multicolumn{4}{|c|}{\textbf{Classifier:} Random Forest}								\\ \hline
    \multicolumn{4}{|c|}{\textbf{Efficiency:} 50\%}											\\ \hline
    \textbf{Party}  & \textbf{Pre}			& \textbf{Recall}	& \textbf{F-measure}		\\ \hline
    AAP      		& 0.483					& 0.412				& 0.44 	 	    			\\ \hline
    BJP           	& 0.439					& 0.486				& 0.462      				\\ \hline
    CONG 			& 0.588				    & 0.606				& 0.597    					\\ \hline
    \end{tabular}
\captionof{table}{User features based classification results with equal instances of all classes in `Pro' category}\label{tab:result10}
\end{minipage}

When we fed equal number of instances to the same classifier, we saw a further improved performance as compared to the hashtag based classification. The other parameters also showed improvement and were approximately close to 0.5. We then compared this method for the 2-class classification to see if the method was good enough even with 2 classes. The 2 pairs of classifications we did were for AAP-BJP and AAP-Congress. The results shown in Tables \ref{tab:result11} and \ref{tab:result12} show marked improvement than the other methods. So it shows that 

\begin{minipage}[t]{.45\linewidth}
  \centering
    \begin{tabular}{|m{1cm}|m{1cm}|m{1cm}|m{1.5cm}|}
    \hline
    \multicolumn{4}{|c|}{\textbf{Instances:} 266}											\\ \hline
    \multicolumn{4}{|c|}{\textbf{Attributes:} 9}											\\ \hline
    \multicolumn{4}{|c|}{\textbf{Classifier:} Random Forest}								\\ \hline
    \multicolumn{4}{|c|}{\textbf{Efficiency:} 61.27\%}										\\ \hline
	\textbf{Party}  & \textbf{Pre}	& \textbf{Recall}	& \textbf{F-measure}				\\ \hline
    AAP      		& 0.606					& 0.62				& 0.615 	     			\\ \hline
    BJP           	& 0.781					& 0.59  			& 0.603      				\\ \hline
    \end{tabular}
\captionof{table}{User features based classification for AAP-BJP}\label{tab:result11}
\end{minipage}
\hspace{5mm}
\begin{minipage}[t]{.45\linewidth}
  \centering
    \begin{tabular}{|m{1cm}|m{1cm}|m{1cm}| m{1.5cm}|}
    \hline
    \multicolumn{4}{|c|}{\textbf{Instances:} 66}											\\ \hline
    \multicolumn{4}{|c|}{\textbf{Attributes:} 9}											\\ \hline
    \multicolumn{4}{|c|}{\textbf{Classifier:} Random Forest}								\\ \hline
    \multicolumn{4}{|c|}{\textbf{Efficiency:} 71.21\%}											\\ \hline
    \textbf{Party}  & \textbf{Pre}			& \textbf{Recall}	& \textbf{F-measure}		\\ \hline
    AAP      		& 0.706					& 0.727				& 0.716	 	    			\\ \hline
    CONG 			& 0.719				    & 0.697				& 0.708    					\\ \hline
    \end{tabular}
\captionof{table}{User features based classification with AAP-Congress}\label{tab:result12}
\end{minipage}

We did similar experiments for the `Anti' category as well, but the results in that case showed some decline in the efficiency score when fed with all instances, which was 47.76\%. However, when the classifier was provided with equal number of instances of `Anti' category, the efficiency improved by 6\%, making it 41.17\% as compared to the results of hashtag based method. Though the method showed improvement for the `Pro' category, we wanted to see if there was a possible method by which we could achieve further improvement in all 3 parameters. To see that we tried the method which has been explained in the next section.

\section{Network Based Classification}
In this method, we exploited yet another feature of twitter, i.e. the retweet and user mention network. User mentions and retweets are important parts of the tweets composed by the users. When a user tags another user in his tweet, he can do so by using the `@' symbol followed by the user handle \cite{honey2009beyond}. Similarly, one can retweet the tweet posted by other user to broadcast his tweets \cite{boyd2010tweet}. It has been observed that users usually retweet the tweets they like and endorse, though it is not true in every case. Many people are seen mentioning the leaders they follow and retweeting their tweets. This gives us an opportunity to explore if this network of retweets and user\_mentions can help us in our classification. Both these features can be obtained from the JSON object of the tweet returned by the API. The retweeted\_status and user\_mentions object in the entities gives us this information. 

To create the network, we used the retweet and user mentions. First to look at the retweets, each retweet created two nodes. One who retweeted and the other as the one whose tweet has been retweeted. For e.g., user A retweeted or re-posted user B's tweet, then both A and B will be two nodes in the graph and there would be an edge between them. Since the edge is undirected, the direction does not matter, but for the sake of clarity, the edge would go from A to B. Similarly, for the user mentions, if user A mentions B, then an edge between node A and B is created. Please note that if the user A retweets or mentions user B more than once, there would still be just 1 edge between them and in case user B retweets or mentions user B, even then there would not be any extra edge. Once this network was formed, we put this into Gephi 0.8.2, a network analysis and visualization tool.\footnote{\url{http://en.wikipedia.org/wiki/Gephi}} We then ran the community detection algorithm on this network to develop different clusters in this network. The algorithm used in Gephi is Fast unfolding of communities in large networks by Blondel et. al.\cite{blondel2008fast}. This algorithm is claimed to be the fastest for large networks by the authors. The modularity score is the measure of how well are the communities separated in the network \cite{newman2006finding}. Once the communities were formed, we used the force atlas \cite{fruchterman1991graph} layout in Gephi to display the communities in the network. The exact community of each node was then stored in a CSV (Comma Separated Values) file. We looked at the communities and decided them to be belonging to any one party based on the screen\_names appearing in them. If the community had maximum profiles belonging to BJP, then it was named BJP. We then checked the orientation assigned by the community detection algorithm to the actual orientation of the user and calculated the efficiency. The results have been discussed below.

\begin{figure*}
  \centering
  \subfigure[]{%
    \includegraphics[width=0.45\textwidth]{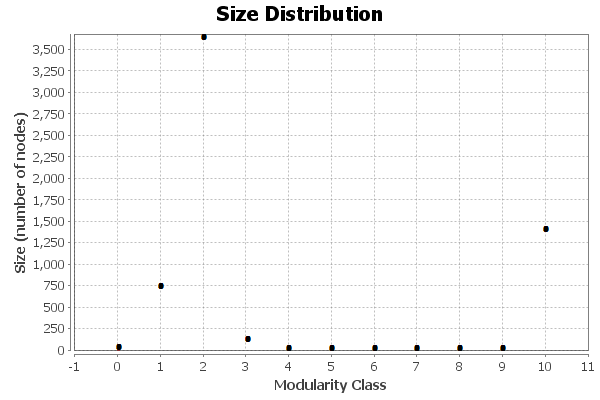}%
    \label{fig:distribution}%
  }%
  \hfill
  \subfigure[]{%
    \includegraphics[width=0.45\textwidth]{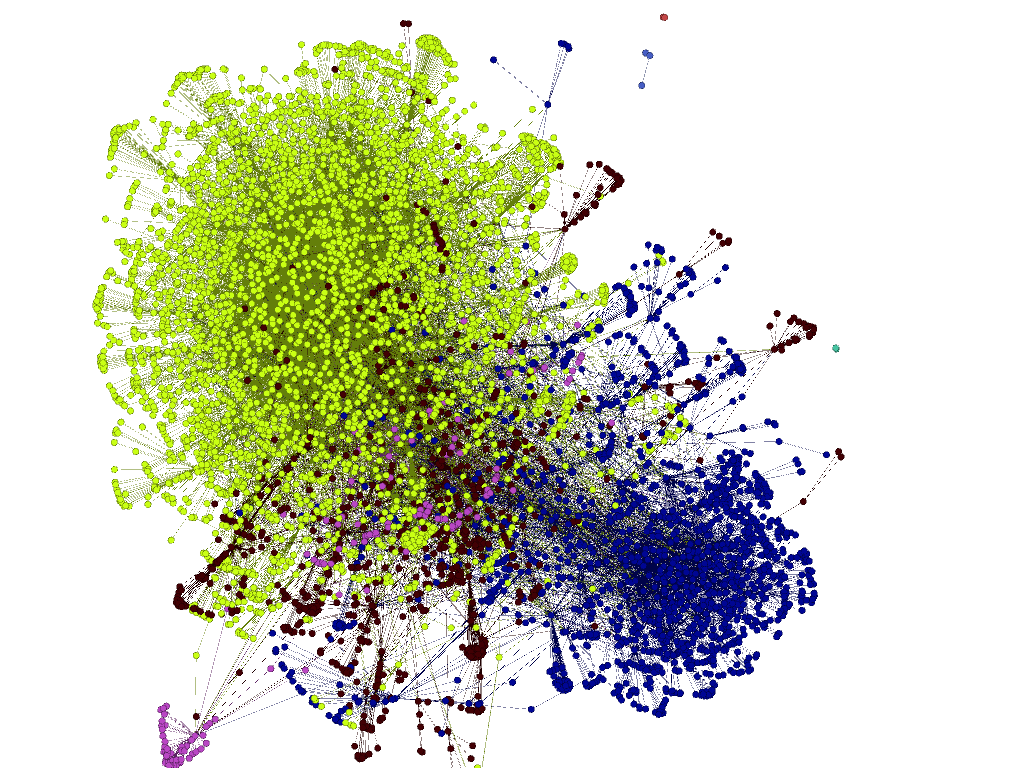}%
    \label{fig:network1}%
  }%
   \caption[Network formed with all 613 `Pro' instances ]{(a) Size distribution of the nodes for all 613 instances (b) Force directed graph of retweet and user mentions network with all 613 instances }
  \label{fig:networkall}
\end{figure*}

As in the previous cases, we took up all the 613 instances. The retweets and user mentions gave us 6,022 nodes and 13,693 edges. After running the community detection algorithm, . The algorithm resulted in formation of 11 communities, 3 of which were the major ones numbered 2, 10 and 1 with 62.27\%, 23.24\% and 12.45\% of the nodes respectively. The rest of the communities had less than 0.05\% of the nodes and we thus chose to ignore them. The modularity score for this case was 0.402. The size distribution can be seen in Figure \ref{fig:distribution}. The network we got as the force atlas layout is shown in Figure \ref{fig:network1}

Upon scanning the clusters, we found that the one with green color had most of the BJP related accounts, so we gave it the tag of BJP, the dark blue had maximum related to AAP and was thus assigned as AAP and the dark brown ones was Congress. The result of this analysis gave us an efficiency of 78.35\%. The precision  and recall values are shown in the Table \ref{tab:result13}.

\begin{minipage}[t]{.45\linewidth}
  \centering
    \begin{tabular}{|m{1cm}|m{1cm}|m{1cm}|m{1.5cm}|}
    \hline
    \multicolumn{4}{|c|}{\textbf{\#Nodes:} 6022}											\\ \hline
    \multicolumn{4}{|c|}{\textbf{\#Edges:} 13693}											\\ \hline
    \multicolumn{4}{|c|}{\textbf{Modularity Score:} 0.402}									\\ \hline
    \multicolumn{4}{|c|}{\textbf{Efficiency:} 78.31\%}										\\ \hline
	\textbf{Party}  & \textbf{Pre}	& \textbf{Recall}	& \textbf{F-measure}				\\ \hline
    AAP      		& 0.709					& 0.672				& 0.690 	     			\\ \hline
    BJP           	& 0.939					& 0.850  			& 0.897      				\\ \hline
    CONG           	& 0.326					& 0.576  			& 0.451      				\\ \hline
    \end{tabular}
\captionof{table}{Results for network based classification for 613 instances}\label{tab:result13}
\end{minipage}
\hspace{5mm}
\begin{minipage}[t]{.45\linewidth}
  \centering
    \begin{tabular}{|m{1cm}|m{1cm}|m{1cm}| m{1.5cm}|}
    \hline
    \multicolumn{4}{|c|}{\textbf{\#Nodes:} 1193}											\\ \hline
    \multicolumn{4}{|c|}{\textbf{\#Edges:} 1489}											\\ \hline
    \multicolumn{4}{|c|}{\textbf{Modularity Score:} 0.582}									\\ \hline
    \multicolumn{4}{|c|}{\textbf{Efficiency:} 80.00\%}										\\ \hline
	\textbf{Party}  & \textbf{Pre}	& \textbf{Recall}	& \textbf{F-measure}				\\ \hline
    AAP      		& 0.856					& 0.733				& 0.794 	     			\\ \hline
    BJP           	& 0.769					& 0.952  			& 0.860      				\\ \hline
    CONG           	& 0.818					& 0.897  			& 0.857      				\\ \hline
    \end{tabular}
\captionof{table}{Results for network based classification for equal instances of `Pro' category}\label{tab:result14}
\end{minipage}

\begin{figure}[!ht]
\centering
\includegraphics[scale=.4]{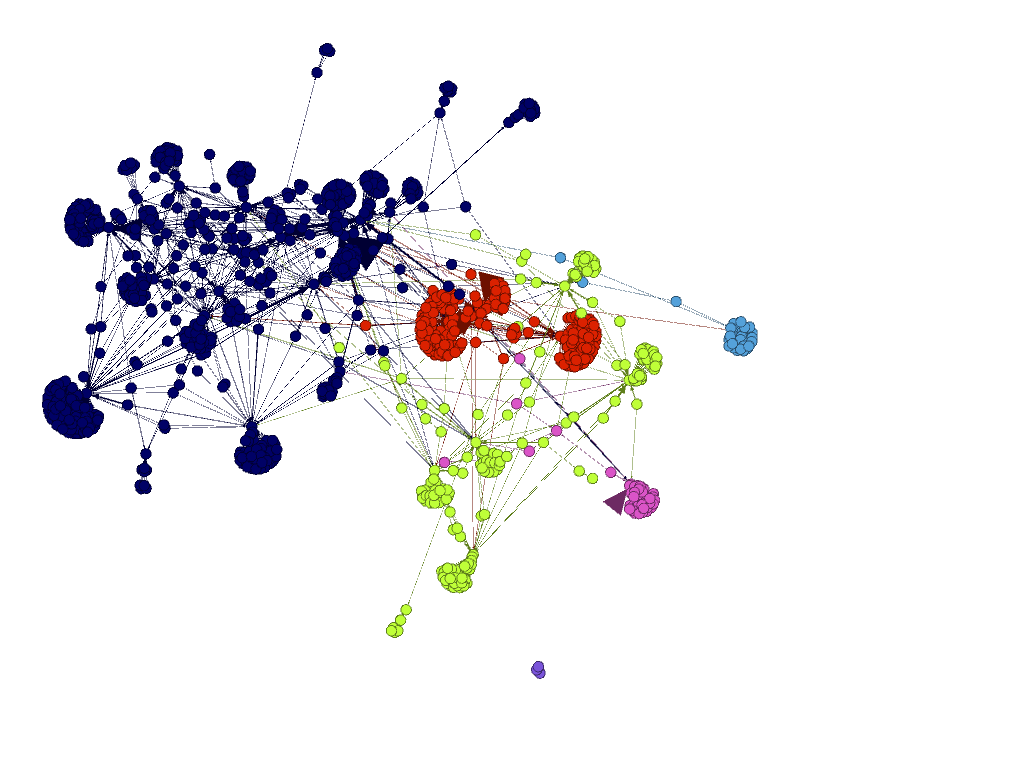}
\caption[Network formed from equal instances of all classes]{Network formed from equal instances of all classes}
\label{fig:networkequal}
\end{figure}

Just to remove all kinds of bias from the data, we took an equal sample of all the classes and drew a network for the 99 instances. The network of 1193 nodes resulted in 8 communities with 5 of them having 0.05\% nodes. The most densely populated were community number 2 with 56.67\% nodes and most of them belonging to BJP accounts; the next being community 0 with 18.44\% nodes and majorly formed by accounts supporting AAP. Similarly the third largest community was the one with 16.51\% belonging to Congress. The network formed is shown in Figure \ref{fig:networkequal}. The dark blue colored cluster is the cluster of BJP users, the green is that of Congress, whereas the red ones depict AAP users. The results show 80\% efficiency for this method with equal number of instances, the highest till now.

We did a 2-class classification as well for this method and recorded the highest percentage of efficiency with 87.79\%. This was the efficiency for the pair of AAP and BJP. Similar higher efficiency for other pairs was also observed making the network analysis method the most suitable for classification. 

\chapter{System Design}
In this chapter we shall discuss the system we built for the daily analysis of tweets. In each of the sections we describe the tabs of the portal. The portal address is
\begin{center}
\url{http://bheem.iiitd.edu.in/IndiaElections}
\end{center}

\section{Real-Time Tweets}
\begin{figure}[!ht]
\centering
\includegraphics[scale=.35]{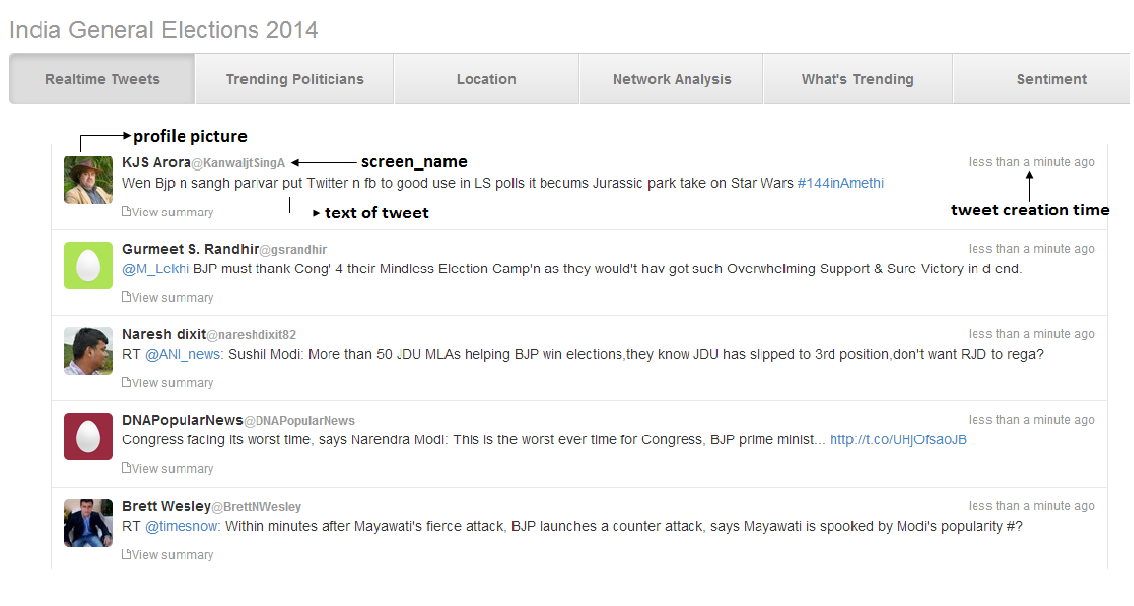}
\caption[Screen shot of the `Realtime Tweets' tab of the portal]{Screen shot of the `Realtime Tweets' tab of the portal showing various fields of a tweet}
\label{fig:ss1}
\end{figure}

As stated in the Section \ref{3.2.1}, we wanted to have a system where the tweets related to elections could be seen as and when they came. So we displayed all the real time tweets on this tab. The tweets here were shown in nearly the same manner as in Twitter. The profile picture of the person tweeting and the time of creation of tweet was also displayed along with the text. Figure \ref{fig:ss1} points the various fields of the tweet such as the text of the tweet, the display picture, the time of creation of the tweet.

Since we wanted the page to be completely real time, we wanted to display the new tweets as well. For this we made the Javascript to look for new tweets, if any, at an interval of 5000 milliseconds. If there were any new tweets, a ticker or a status bar would show up on the page, displaying the number of newly found tweets. If the user clicked on the ticker, the new tweets would show up on the top, pushing the old ones to bottom. 

\section{Trending Politicians}\label{trending politicians}
For the `Trending Politicians' tab, we displayed a table with twitter accounts of all the prominent national leaders and parties which were picked by us. Most of these accounts were verified and legitimate. The content of this table was generated from the data collected (see Section \ref{3.1.2}) at an interval of 24 hours for all the 130 verified accounts. 

\begin{figure}[!ht]
\centering
\includegraphics[scale=.5]{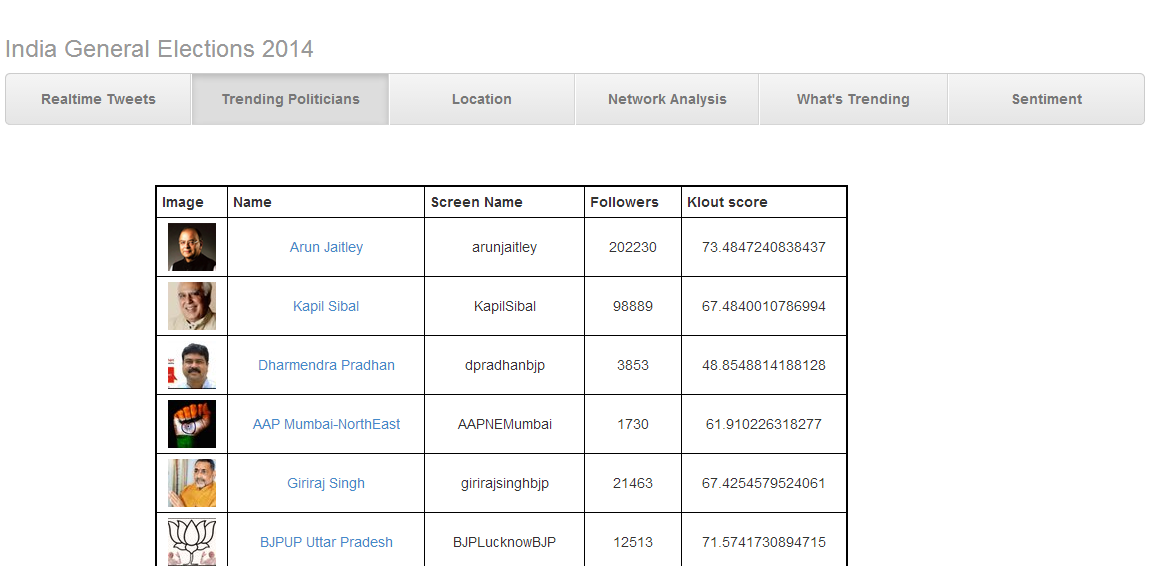}
\caption[Screen shot of the `Trending Politicians' tab]{Screen shot of the `Trending Politicians' tab of the portal}
\label{fig:ss3}
\end{figure}

Figure \ref{fig:ss3} shows the table in the `Trending Politicians' tab. The table in this tab has 5 columns: first column has the display picture of the user as a thumbnail. The next column displays the name being used the user. The third column contains the screen name of the user. This is the name used to mention the user in tweets with a `@' symbol. The third column is clickable and contains the links to the twitter profile of the users. So if one wants to have a look at the profile or tweets of that user, he can click on the screen\_name in the third column and he would be redirected to the twitter profile of that user. The fourth column mentions the number of followers of the users. The last column displays the Klout score of the user.

Klout score is a value between 1-100 that represents the influence score of the  user.\footnote{\url{https://klout.com/corp/score}} We made use of the KloutAPI v2 to calculate the Klout score.\footnote{\url{http://klout.com/s/developers/home}} This score is calculated for each user, taking into consideration factors such as the number of friends, followers, list memberships, how many spam accounts are following the user etc. A higher Klout score indicates higher influence of the user on the social media network. This score was thus required to see which politicians or parties were having a higher influence on the social media.

The table in this tab can also be sorted on the number of followers and the Klout score in both increasing and decreasing orders. This feature was included so that one can easily see the profiles that were having maximum/ minimum number of followers or influence factor. 

\section{Location of Tweets}
\begin{figure}[!ht]
\centering
\includegraphics[scale=.5]{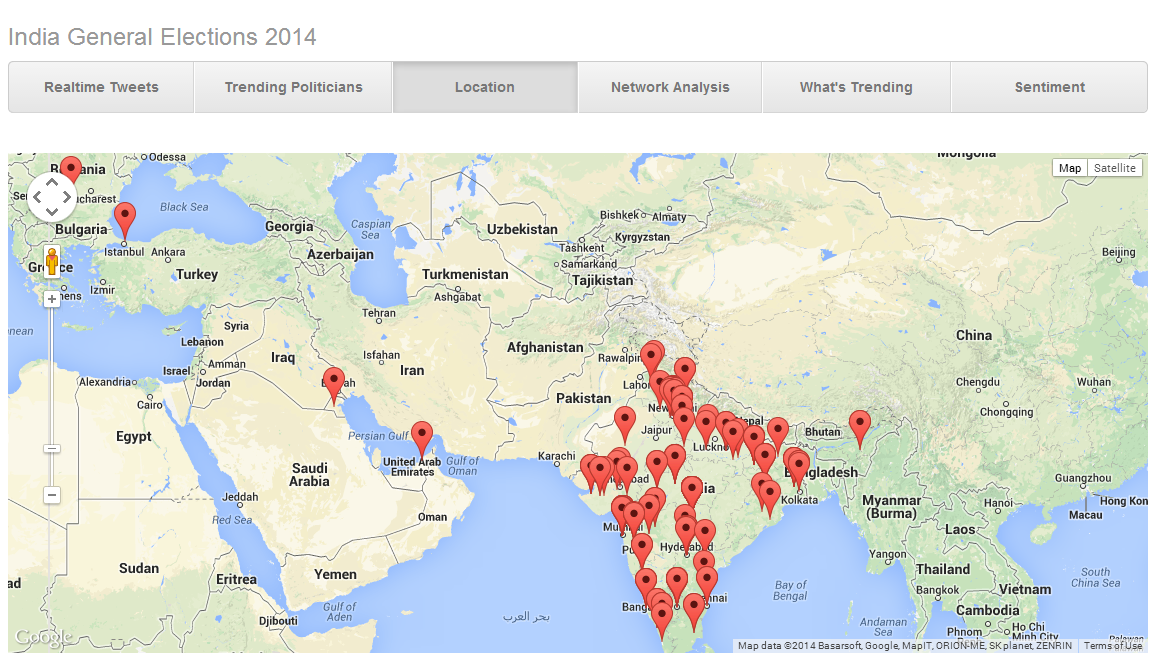}
\caption[`Location' tab of the portal]{Screen shot of the `Location' tab showing the markers at the location of the tweets}
\label{fig:ss5}
\end{figure}

In this section, we shall discuss the `Location' tab of the portal. In this tab we used the Google Maps to visualize the location from where the tweets were coming. To find the location of the tweet, we first checked if the tweet was geotagged. If yes, then the location, specified by the latitude and longitude was used to create a marker at the particular position in the map. If the tweet was not geotagged, we used the location of the user as the location of the tweet. The location of the user is a string. We access the geocoding service in the Google Maps API service with the Geocoder.geocode() method within our code, to obtain the latitude and longitude values from the location string. This service tries to find the best match for the string. We then put markers on these lat long values on the map with the help of createMarker() method. The result can be seen in Figure \ref{fig:ss5} We chose to display the location for 300 tweets on the map.

\section{Network Analysis}
\begin{figure}[!ht]
\centering
\includegraphics[scale=.5]{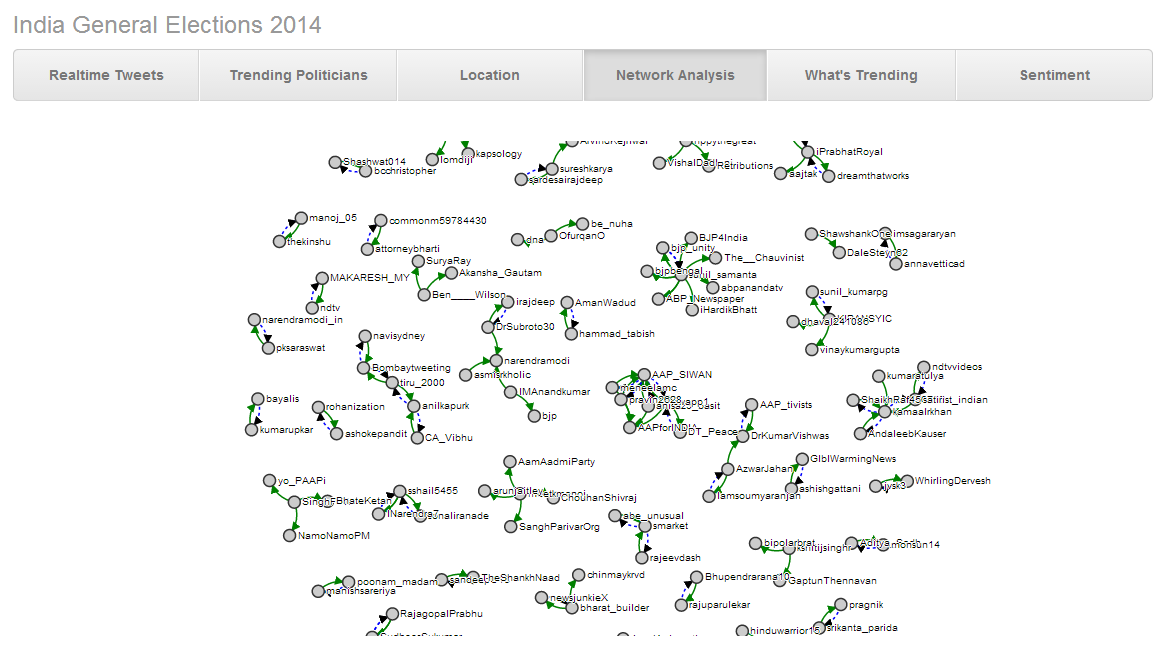}
\caption[Screen shot `Network Analysis' tab of the portal]{Screen shot of the `Network Analysis' tab}
\label{fig:ss6}
\end{figure}

In the `Network Analysis' tab of the portal, we put up a force directed network graph of the user mentions and retweets of the tweets. In the tweets, if a user wants to mention other user then he makes use of his screen\_name appended by a `@' symbol. Such mentions are called the user mentions and can be obtained from  JSON object of the tweet returned by the API. Similarly, if the tweet is a retweet, the JSON object returned would have a field called retweeted\_status, which would contain the screen\_name of the user being retweeted. In Figure \ref{fig:ss6}, we see the graph where the green color arrows represent the user\_mentions, whereas the blue dotted arrows represent the retweet.

\section{What's Trending}
For the `What's Trending' tab, we came up with a word cloud and a graph showing the per hour frequency of the top 10 hashtags. To generate the word cloud, we pick up the tweets from last 24 hours and use the text field of the tweets. We then remove all the stop words, user\_mentions, hashtags and URLs from the text of the tweet. Then we calculate the frequency of each word in all the tweets over the last 24 hours. The top 50 most frequently occurring words are chosen to be displayed in the word cloud. This word cloud is generated with d3.json template. The size of each word is proportional to its frequency of occurrence. The color and position of the word is randomly chosen.

\begin{figure}[!ht]
\centering
\includegraphics[scale=.5]{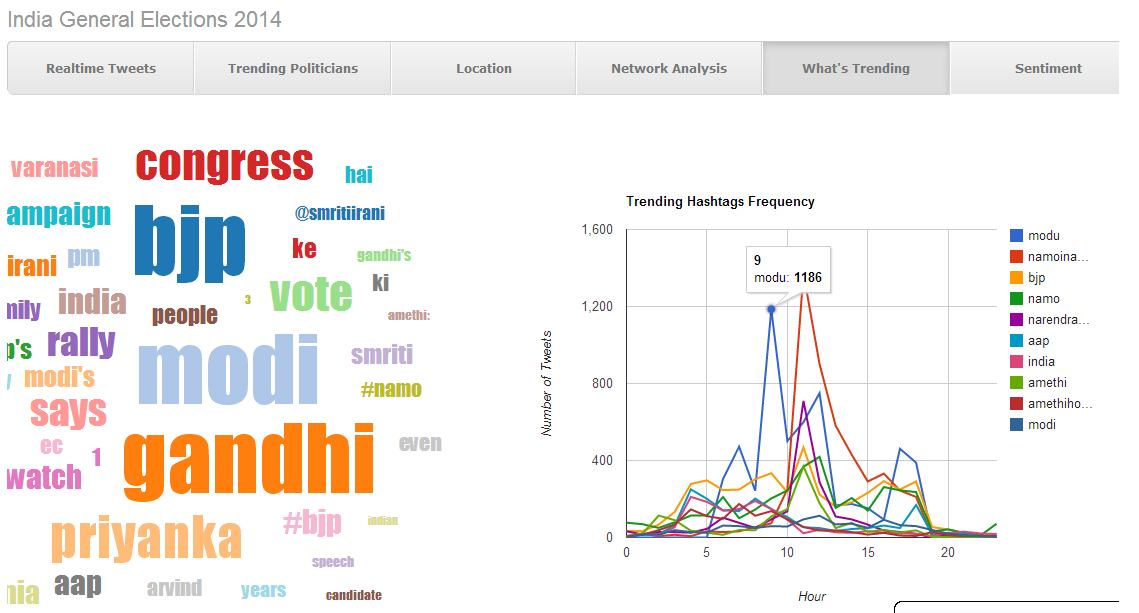}
\caption[`What's Trending' tab of the portal]{Screen shot of the `What's Trending' tab showing the word cloud on the left and hashtag per hour frequency graph on the right}
\label{fig:ss7}
\end{figure}

For the part of this tab, we do an analysis on the per hour frequency of the hashtags. Since hashtags give us an idea about the topic of the tweet, it was necessary to see which topics were trending at what hour of the day in the last 24 hours. To this effect, we did a volume count of the hashtags occurring in all the tweets and then summed them over hours. The top 10 with highest frequency were chosen and displayed on the graph generated with the Google Charts API.\footnote{\url{https://developers.google.com/chart/interactive/docs/gallery/linechart}} Figure \ref{fig:ss7} shows the word cloud and the hashtag frequency graph.

One can click on the words in the word cloud and at the data points in the hashtag frequency graph to see the tweets that contained those words or hashtags. Those tweets would be displayed in a pop up modal window. One can also click on the profile picture or the screen name of the users who tweeted to go to their specific profile. 

\section{Sentiment Analysis}
\begin{figure*}
  \centering
  \subfigure[]{%
    \includegraphics[scale=.4]{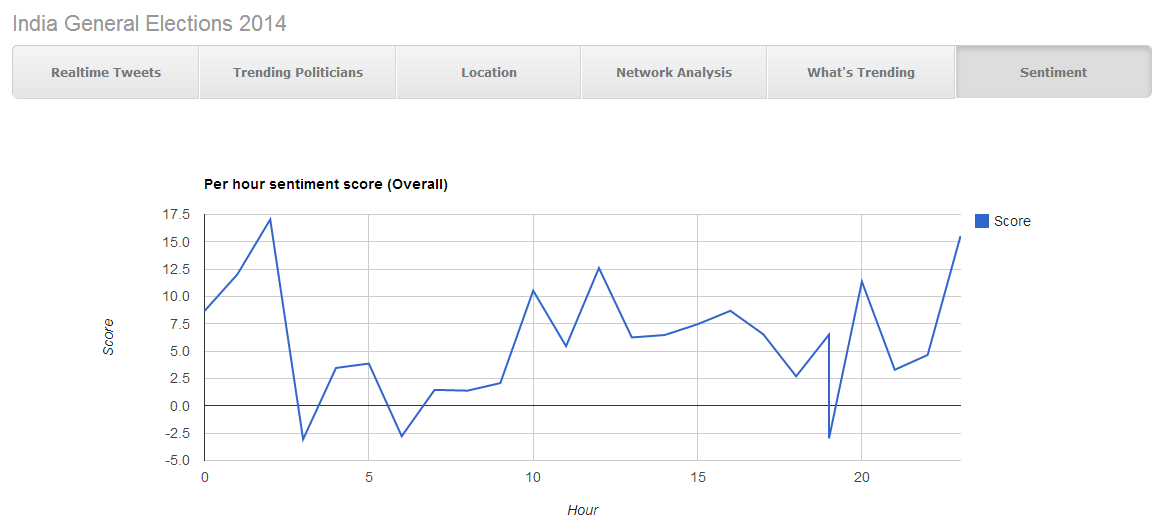}%
    \label{fig:ss9}%
  }%
  \hfill
  \subfigure[]{%
    \includegraphics[scale=.4]{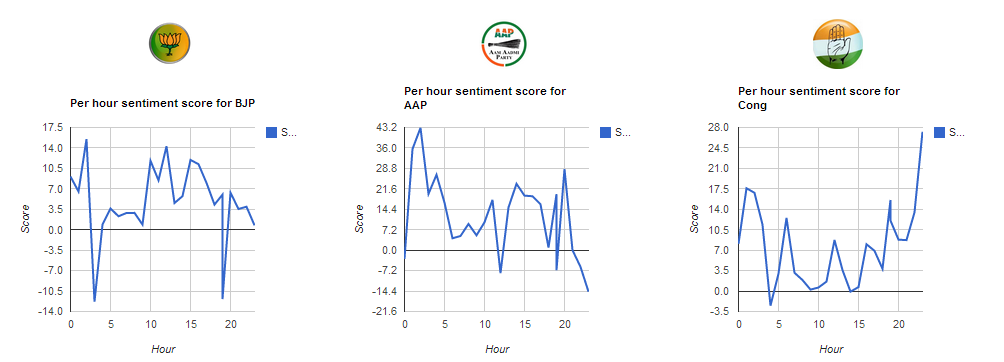}%
    \label{fig:ss10}%
  }%
   \caption[Screen shot of the `Sentiments' tab]{(a) The average sentiment score for every hour for all the tweets (b) The average sentiment score over the past 24 hours for AAP, BJP and Congress }
  \label{fig:senti}
\end{figure*}

The last tab of the portal displayed the sentiments of the tweets. The sentiments for the tweets were calculated with the help of Sentiment140 API.\footnote{\url{http://www.sentiment140.com/api}} We have 4 graphs in this tab. The first one displays the cumulative score for sentiments for all the tweets over the past 24 hours. The next 3 graphs display the sentiment score for the tweets related to individual parties only. For e.g., the graph for AAP has the average sentiment score for every hour for the tweets related to AAP only. So from this tab, we can see the rise or fall in the sentiments of the tweets over the hours. Figures \ref{fig:ss9} and \ref{fig:ss10} show these graphs.

\chapter{Conclusions, Limitations and Future Work}
\section{Conclusions}
In the first part of our work, we analyzed the 17.60 million tweets in our dataset. We analyzed the volume of the tweets incoming every day and justified the peaks in the data by providing the timeline of the major political activities during India General Elections 2014. It was reported that Tuesdays and Wednesdays saw the highest number of Tweets as majority of the activities were during weekdays and the activity peaked particularly in the second half of the day. We also found out that we had 815,425 unique users in our dataset. The number of unique users for every month were also reported. It was shown that as per the popular belief the volume of tweets increased as elections came closer. The number of mentions over past 4 months were shown for not only AAP, BJP and Congress, but 8 other parties that were active in various states. We also analyzed the popularity of Arvind Kejriwal and Narendra Modi on two different parameters and reported how their political behavior affected their popularity on Twitter.

The second part of the work was concerning the political orientation of users. We used 4 approaches for the developing 4 types of classifiers. After getting 1000 profiles annotated we could prepare a true positive dataset with 613 instances of Pro and 425 instance of Anti category. The first approach used a user vector that had the TFIDF score for each term. The efficiency with this method for equal number of instances was 42.42\% for Pro and 37.25\% for the Anti category. We then moved on to the second method which used the hashtags used in tweets as features for classification. This method improved the efficiency to some extent but not to acceptable limits. To verify if the methods were working right, we tried 2-class classification as well, but to no avail. The third method was the use of user based features such as number of friends and followers, number of occurrences of AAP, BJP and Congress related words and hashtags . This method increased the efficiency above 50\% for the Pro category and over 40\% for the Anti category. The last method we tried was the community detection algorithm on the retweet and user mentions network, which gave us an efficiency of over 80\%. So we can conclude from our study that analyzing the political alignment of Indian users with the content based techniques is a hard problem and one can get success only with the network analysis. Also the classification of Anti political alignment of users is further more difficult than the Pro category classification.

The third and last part of the work was the development of the portal that displayed the analysis of the tweets of the last 24 hours. We displayed the real time tweets on our portal as and when they came. We also kept a daily track of the timeline of the political users and maintained their Klout score as well. Our portal showed the network analysis and locations of the tweets coming in. We segregated the tweets as belonging to AAP, BJP and Congress and then analyzed the positive and negative sentiments on them. The trending hashtags and words were displayed as tagcloud and frequency graph on our portal.

\section{Limitations}
Since the data we collected was from Twitter API, which gives a limited access to the actual data, we do not know for sure if the data we have is really representative of the complete data. Most of the tweets are not geo-tagged and thus doing a location analysis on them is not very conclusive. Since the languages used in the tweets here are too varied, one can not easily detect the sentiment in them. Too much of data `Pro' to BJP and excess number of parties and candidates made the classification task tough for us.

\section{Future Work}
One can try classifying the gender of the users tweeting in the dataset and try to compare it with the actual demographics. A correlation between the vote share and tweet share can be reported as a part of future work. One can also see how have the sentiments changed after the declaration of results. One can also analyze all the users tweeting about elections to be bots or humans.
\bibliographystyle{acm}
\bibliography{elections}

\begin{thebibliography}{10}

\bibitem{aramaki2011twitter}
{\sc Aramaki, E., Maskawa, S., and Morita, M.}
\newblock Twitter catches the flu: detecting influenza epidemics using twitter.
\newblock In {\em Proceedings of the Conference on Empirical Methods in Natural
  Language Processing\/} (2011), Association for Computational Linguistics,
  pp.~1568--1576.

\bibitem{blondel2008fast}
{\sc Blondel, V.~D., Guillaume, J.-L., Lambiotte, R., and Lefebvre, E.}
\newblock Fast unfolding of communities in large networks.
\newblock {\em Journal of Statistical Mechanics: Theory and Experiment 2008},
  10 (2008), P10008.

\bibitem{bollen2011modeling}
{\sc Bollen, J., Mao, H., and Pepe, A.}
\newblock Modeling public mood and emotion: Twitter sentiment and
  socio-economic phenomena.
\newblock In {\em ICWSM\/} (2011).

\bibitem{boyd2010tweet}
{\sc Boyd, D., Golder, S., and Lotan, G.}
\newblock Tweet, tweet, retweet: Conversational aspects of retweeting on
  twitter.
\newblock In {\em System Sciences (HICSS), 2010 43rd Hawaii International
  Conference on\/} (2010), IEEE, pp.~1--10.

\bibitem{bruns2011use}
{\sc Bruns, A., and Burgess, J.~E.}
\newblock The use of twitter hashtags in the formation of ad hoc publics.

\bibitem{carletta1996assessing}
{\sc Carletta, J.}
\newblock Assessing agreement on classification tasks: the kappa statistic.
\newblock {\em Computational linguistics 22}, 2 (1996), 249--254.

\bibitem{castillo2011information}
{\sc Castillo, C., Mendoza, M., and Poblete, B.}
\newblock Information credibility on twitter.
\newblock In {\em Proceedings of the 20th international conference on World
  wide web\/} (2011), ACM, pp.~675--684.

\bibitem{cohen2013classifying}
{\sc Cohen, R., and Ruths, D.}
\newblock Classifying political orientation on twitter: It’s not easy!
\newblock In {\em Proceedings of the 7th International Conference on Weblogs
  and Social Media\/} (2013).

\bibitem{conover2011predicting}
{\sc Conover, M.~D., Gon{\c{c}}alves, B., Ratkiewicz, J., Flammini, A., and
  Menczer, F.}
\newblock Predicting the political alignment of twitter users.
\newblock In {\em Privacy, security, risk and trust (passat), 2011 ieee third
  international conference on and 2011 ieee third international conference on
  social computing (socialcom)\/} (2011), IEEE, pp.~192--199.

\bibitem{fruchterman1991graph}
{\sc Fruchterman, T.~M., and Reingold, E.~M.}
\newblock Graph drawing by force-directed placement.
\newblock {\em Software: Practice and experience 21}, 11 (1991), 1129--1164.

\bibitem{gayo2011don}
{\sc Gayo-Avello, D.}
\newblock Don't turn social media into another'literary digest'poll.
\newblock {\em Communications of the ACM 54}, 10 (2011), 121--128.

\bibitem{golbeck2011computing}
{\sc Golbeck, J., and Hansen, D.}
\newblock Computing political preference among twitter followers.
\newblock In {\em Proceedings of the SIGCHI Conference on Human Factors in
  Computing Systems\/} (2011), ACM, pp.~1105--1108.

\bibitem{gupta2012credibility}
{\sc Gupta, A., and Kumaraguru, P.}
\newblock Credibility ranking of tweets during high impact events.
\newblock In {\em Proceedings of the 1st Workshop on Privacy and Security in
  Online Social Media\/} (2012), ACM, p.~2.

\bibitem{himelboim2013birds}
{\sc Himelboim, I., McCreery, S., and Smith, M.}
\newblock Birds of a feather tweet together: Integrating network and content
  analyses to examine cross-ideology exposure on twitter.
\newblock {\em Journal of Computer-Mediated Communication 18}, 2 (2013),
  40--60.

\bibitem{honey2009beyond}
{\sc Honey, C., and Herring, S.~C.}
\newblock Beyond microblogging: Conversation and collaboration via twitter.
\newblock In {\em System Sciences, 2009. HICSS'09. 42nd Hawaii International
  Conference on\/} (2009), IEEE, pp.~1--10.

\bibitem{hong2011does}
{\sc Hong, S., and Nadler, D.}
\newblock Does the early bird move the polls?: the use of the social media
  tool'twitter'by us politicians and its impact on public opinion.
\newblock In {\em Proceedings of the 12th Annual International Digital
  Government Research Conference: Digital Government Innovation in Challenging
  Times\/} (2011), ACM, pp.~182--186.

\bibitem{jungherr2012pirate}
{\sc Jungherr, A., J{\"u}rgens, P., and Schoen, H.}
\newblock Why the pirate party won the german election of 2009 or the trouble
  with predictions: A response to tumasjan, a., sprenger, to, sander, pg, \&
  welpe, im “predicting elections with twitter: What 140 characters reveal
  about political sentiment”.
\newblock {\em Social Science Computer Review 30}, 2 (2012), 229--234.

\bibitem{krenn2004determining}
{\sc Krenn, B., Evert, S., and Zinsmeister, H.}
\newblock Determining intercoder agreement for a collocation identification
  task.
\newblock In {\em Proceedings of KONVENS\/} (2004), pp.~89--96.

\bibitem{metaxas2011not}
{\sc Metaxas, P.~T., Mustafaraj, E., and Gayo-Avello, D.}
\newblock How (not) to predict elections.
\newblock In {\em Privacy, security, risk and trust (PASSAT), 2011 IEEE third
  international conference on and 2011 IEEE third international conference on
  social computing (SocialCom)\/} (2011), IEEE, pp.~165--171.

\bibitem{morstatter2013sample}
{\sc Morstatter, F., Pfeffer, J., Liu, H., and Carley, K.~M.}
\newblock Is the sample good enough? comparing data from twitter’s streaming
  api with twitter’s firehose.
\newblock {\em Proceedings of ICWSM\/} (2013).

\bibitem{mustafaraj2011vocal}
{\sc Mustafaraj, E., Finn, S., Whitlock, C., and Metaxas, P.~T.}
\newblock Vocal minority versus silent majority: Discovering the opionions of
  the long tail.
\newblock In {\em Privacy, security, risk and trust (passat), 2011 ieee third
  international conference on and 2011 ieee third international conference on
  social computing (socialcom)\/} (2011), IEEE, pp.~103--110.

\bibitem{newman2006finding}
{\sc Newman, M.~E.}
\newblock Finding community structure in networks using the eigenvectors of
  matrices.
\newblock {\em Physical review E 74}, 3 (2006), 036104.

\bibitem{nowak1990private}
{\sc Nowak, A., Szamrej, J., and Latan{\'e}, B.}
\newblock From private attitude to public opinion: A dynamic theory of social
  impact.
\newblock {\em Psychological Review 97}, 3 (1990), 362.

\bibitem{sakaki2010earthquake}
{\sc Sakaki, T., Okazaki, M., and Matsuo, Y.}
\newblock Earthquake shakes twitter users: real-time event detection by social
  sensors.
\newblock In {\em Proceedings of the 19th international conference on World
  wide web\/} (2010), ACM, pp.~851--860.

\bibitem{skoric2012tweets}
{\sc Skoric, M., Poor, N., Achananuparp, P., Lim, E.-P., and Jiang, J.}
\newblock Tweets and votes: A study of the 2011 singapore general election.
\newblock In {\em System Science (HICSS), 2012 45th Hawaii International
  Conference on\/} (2012), IEEE, pp.~2583--2591.

\bibitem{tumasjan2010predicting}
{\sc Tumasjan, A., Sprenger, T.~O., Sandner, P.~G., and Welpe, I.~M.}
\newblock Predicting elections with twitter: What 140 characters reveal about
  political sentiment.
\newblock {\em ICWSM 10\/} (2010), 178--185.

\bibitem{tumasjan2011election}
{\sc Tumasjan, A., Sprenger, T.~O., Sandner, P.~G., and Welpe, I.~M.}
\newblock Election forecasts with twitter how 140 characters reflect the
  political landscape.
\newblock {\em Social Science Computer Review 29}, 4 (2011), 402--418.

\bibitem{wu2008interpreting}
{\sc Wu, H.~C., Luk, R. W.~P., Wong, K.~F., and Kwok, K.~L.}
\newblock Interpreting tf-idf term weights as making relevance decisions.
\newblock {\em ACM Transactions on Information Systems (TOIS) 26}, 3 (2008),
  13.

\end{thebibliography}


\end{document}